\RequirePackage[svgnames]{xcolor}

\documentclass[11pt,letterpaper]{mystyle}

\usepackage[all]{hypcap}
\usepackage[svgnames]{xcolor}
\usepackage[comma,authoryear,compress]{natbib}
\usepackage{xspace}

\usepackage{hyperref}[citecolor=lightblue]

\hypersetup{
    colorlinks = true,
    citecolor = {YaleBlue},
}

\usepackage{algorithm}
\usepackage{algorithmicx}
\usepackage{algpseudocode}
\usepackage{microtype}
\usepackage{graphicx}
\expandafter\def\csname ver@subfig.sty\endcsname{}
\usepackage{booktabs} %
\usepackage{float}
\usepackage{bigstrut}

\usepackage{amsmath}
\usepackage{amssymb}
\usepackage{mathtools}
\usepackage{amsthm}
\usepackage{mathrsfs}
\usepackage{nicefrac}
\usepackage{dsfont}
\usepackage{enumitem}
\usepackage{subcaption}
\usepackage{graphicx,subfig}
\usepackage{cleveref}
\usepackage{bxcoloremoji}

\usepackage{float}

\usepackage[utf8]{inputenc} %
\usepackage[T1]{fontenc}    %
\usepackage{hyperref}       %
\usepackage{url}            %
\usepackage{booktabs}       %
\usepackage{amsfonts}       %
\usepackage{nicefrac}       %
\usepackage{microtype}      %
\usepackage{graphicx}
\usepackage{subcaption} 
\usepackage{amssymb}
\usepackage{fdsymbol}
\usepackage{wrapfig}
\usepackage{lipsum}
\usepackage{enumitem}
\usepackage{stackengine}
\usepackage[font=small,labelfont=bf]{caption}
\usepackage{color}
\usepackage{adjustbox}

\usepackage{rotating}
\usepackage{makecell}

\newcommand{\x}{\mathbf{x}}

\definecolor{blanchedalmond}{rgb}{1.0, 0.92, 0.8}
\definecolor{carmine}{rgb}{0.59, 0.0, 0.09}
\definecolor{lightblue}{rgb}{0.22,0.45,0.70}%

\renewcommand{\mathbf}{\boldsymbol}

\makeatletter
\def\Ddots{\mathinner{\mkern1mu\raise\p@
\vbox{\kern7\p@\hbox{.}}\mkern2mu
\raise4\p@\hbox{.}\mkern2mu\raise7\p@\hbox{.}\mkern1mu}}
\makeatother
\def \defpart{RAGPart\xspace}
\def \defmask{RAGMask\xspace}

\definecolor{amaranth}{rgb}{0.9, 0.17, 0.31}
\definecolor{antiquebrass}{rgb}{0.8, 0.58, 0.46}
\definecolor{antiquefuchsia}{rgb}{0.57, 0.36, 0.51}
\definecolor{chromeyellow}{rgb}{0.31, 0.47, 0.26}

\newtcolorbox{AIbox}[2][]{aibox,title=#2,#1}
\definecolor{lightblue}{rgb}{0.22,0.45,0.70}%
\definecolor{Gray}{gray}{0.95}
\definecolor{Cornsilk}{rgb}{1.0, 0.97, 0.86}

\usepackage{booktabs,tabularx,array,makecell,multirow,xcolor}

\newcolumntype{Y}{>{\centering\arraybackslash}X}

\tcbset{
  defensebox/.style={
    on line,
    boxrule=0pt,
    arc=2pt,
    colframe=white,
    width=1.2cm,     
    halign=center,
    valign=center,
    left=2pt, right=2pt, top=1pt, bottom=1pt
  }
}

\newtcbox{\good}{defensebox, colback=green!30}
\newtcbox{\bad}{defensebox, colback=red!30}

\usepackage{pifont}
\usepackage{xcolor}
\definecolor{mygreen}{rgb}{0.0, 0.6, 0.0}
\newcommand{\cmark}{\textcolor{mygreen}{\ding{51}}} 
\newcommand{\xmark}{\textcolor{red}{\ding{55}}}   

\usepackage{amsmath}

\usepackage[all]{hypcap}

\title{  RAGPart \& RAGMask: Retrieval-Stage Defenses Against Corpus Poisoning in Retrieval-Augmented Generation
}

\runningtitle{RAGPart \& RAGMask: Retrieval-Stage Defenses Against Corpus Poisoning in Retrieval-Augmented Generation}

\author[1]{
  Pankayaraj Pathmanathan}
\author[1]{
  Michael-Andrei Panaitescu-Liess}
\author[3]{
  Cho-Yu Jason Chiang 
}
\author[1,2]{
  Furong Huang
}

\affil[1]{University of Maryland College Park}
\affil[2]{Capital One}
\affil[3]{Peraton Labs}

\correspondingauthor{Pankayaraj; Email \href{mailto:pan@umd.edu}{pan@umd.edu}}

\begin{document}
\begin{abstract}

    etrieval-Augmented Generation (RAG) has emerged as a promising paradigm to enhance large language models (LLMs) with external knowledge, reducing hallucinations and compensating for outdated information. However, recent studies have exposed a critical vulnerability in RAG pipelines—\emph{corpus poisoning}—where adversaries inject malicious documents into the retrieval corpus to manipulate model outputs. In this work, we propose two complementary retrieval-stage defenses: \textbf{\defpart} and \textbf{\defmask.} Our defenses operate directly on the retriever, making them computationally lightweight and requiring no modification to the generation model. RAGPart leverages the inherent training dynamics of dense retrievers, exploiting document partitioning to mitigate the effect of poisoned points. In contrast, RAGMask identifies suspicious tokens based on significant similarity shifts under targeted token masking. Across two benchmarks, four poisoning strategies, and four state-of-the-art retrievers, our defenses consistently reduce attack success rates while preserving utility under benign conditions. We further introduce an interpretable attack to stress-test our defenses. Our findings highlight the potential and limitations of retrieval-stage defenses, providing practical insights for robust RAG deployments.
\end{abstract}
\maketitle
\vspace{3mm}

\addtocontents{toc}{\protect\setcounter{tocdepth}{-1}}
\section{Introduction}
\label{sec:intoduction}
Large Language Models (LLMs) \citep{openai2024gpt4technicalreport, deepseekai2025deepseekr1incentivizingreasoningcapability} have demonstrated remarkable capabilities in reasoning, problem-solving, and generalization, fueling deployment in real-world domains such as healthcare \citep{article} and finance \citep{Loukas_2023}. Despite these successes, LLMs remain limited by their static training data, resulting in outdated knowledge, hallucinations \citep{Huang_2025}, and gaps in domain-specific expertise.


Retrieval-Augmented Generation (RAG) has recently gained popularity \citep{lewis2021retrievalaugmentedgenerationknowledgeintensivenlp} as a strategy to mitigate these limitations. RAG enhances LLMs by dynamically retrieving external documents relevant to a query from large corpora—such as Wikipedia \citep{thakur2021beirheterogenousbenchmarkzeroshot} or financial reports \citep{10.1145/3184558.3192301}—and augmenting the model's input context. Typically, documents are retrieved based on embedding similarity, computed by traditional methods \citep{SALTON1988513,10.1561/1500000019} or modern dense retrievers \citep{izacard2022unsuperviseddenseinformationretrieval, wang2024multilinguale5textembeddings}.


\begin{figure*}[!htbp]
     \centering
     \includegraphics[width=0.62\textwidth]{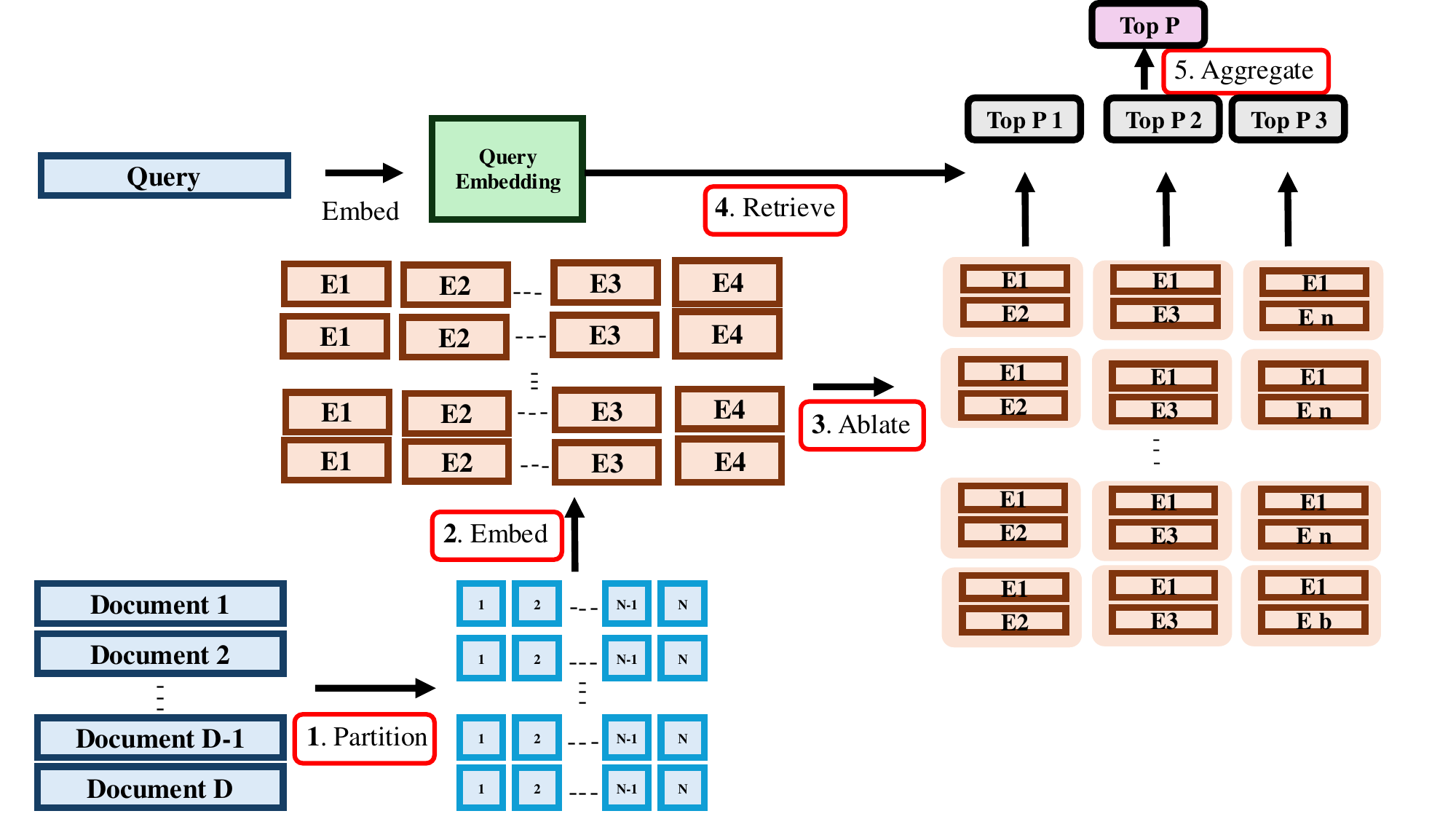} 
     \caption{\textbf{\defpart}: Figure illustrates the \defpart defense, where each document is partitioned into fragments, which are individually embedded. Embeddings from multiple fragment combinations (e.g., subsets of size k) are then averaged to produce candidate document representations. These are used to retrieve multiple top-p document sets, which are aggregated to form the final top-p set for retrieval.}  
     \label{fig:method_1}
\end{figure*}

Despite enhancing factual consistency \citep{Ayala_2024}, RAG's dependence on extensive, publicly sourced corpora introduces vulnerabilities to \emph{corpus poisoning} attacks \citep{zou2024poisonedragknowledgecorruptionattacks}. In these attacks, adversaries insert maliciously crafted documents intended to manipulate the retrieved context, thereby influencing the model's outputs. An attack is successful when a poisoned document is retrieved (retrieval condition) and significantly impacts the LLM's generated response (generation condition).


Current defenses largely focus on the generation stage, proposing certifiable robustness via response aggregation \citep{xiang2024certifiablyrobustragretrieval}. However, these approaches rely on strong and often impractical assumptions. They assume that each retrieved document—often called a \emph{golden document}—is independently sufficient to answer the query, that the retriever can consistently return enough such golden documents, and that it is computationally feasible to run a separate LLM generation for each one. In real-world systems where inference time and cost are major concerns, generating multiple responses per query is typically infeasible, making these defenses difficult to use in practice.

To overcome these issues, practical defenses must satisfy three conditions: \textbf{(W1)} \textit{Effectiveness}: achieving low attack success rates without significant utility loss under benign conditions; \textbf{(W2)} \textit{Efficiency}: remaining computationally lightweight for real-time use; and \textbf{(W3)} \textit{Minimal retriever assumptions}: not requiring perfect or highly accurate retrieval.

To satisfy these properties, we propose addressing corpus poisoning earlier—at the retrieval stage. This is feasible in practice because retrievers are typically smaller models than long-context LLMs, and their similarity computations (e.g., dot products in embedding space) are highly parallelizable. 

Motivated by deep partition-and-aggregation defenses (DPA) \citep{levine2021deeppartitionaggregationprovable, sun2022certifiablyrobustpolicylearning} and perturbation-based defenses like RAP \citep{yang2021raprobustnessawareperturbationsdefending}, we propose two complementary retrieval-stage defenses: \textbf{\defpart} and \textbf{\defmask}. RAGPart leverages dense retrievers' training dynamics, particularly the observation that document fragments often preserve the semantic meaning of the full document in embedding space. For example, dense retrievers such as Contriever \citep{izacard2022unsuperviseddenseinformationretrieval} explicitly define positive training pairs by treating randomly cropped portions of a document as semantically equivalent to the whole, inducing an inductive bias in the retriever's embedding space. We empirically observe that this behavior generalizes across multiple dense retrievers. By exploiting the similarity between full-document and fragment embeddings, \defpart formulates a defense to mitigate the effect of poisoned content. \defmask, on the other hand, targets a different vulnerability: poisoning often hinges on a small set of influential tokens that disproportionately affect similarity scores. By selectively masking these tokens and measuring the resulting similarity shift, \defmask identifies and suppresses poisoned documents. \\

Our contributions are summarized as follows.

\begin{itemize}
    \item 
    We propose two retrieval-stage defenses—\textbf{\defpart} and \textbf{\defmask}—that are both computationally efficient and effective at mitigating corpus poisoning, without modifying the LLM or relying on strong retriever assumptions.
    \item 
    We demonstrate the efficacy of these defenses across two benchmark datasets and four distinct poisoning strategies. In addition to evaluating against existing attacks, we introduce a stronger, interpretable poisoning attack---AdvRAGgen---and show that our defenses remain robust under this more challenging threat model. 
    \item We present a theoretical result demonstrating the superiority of RAGPart over a naive combinatorial approach that does not exploit the properties of dense retrievers.
    \item 
    We systematically analyze trade-offs between \defpart and \defmask in terms of defense effectiveness and computational efficiency, providing practical guidance for real-world deployment across various system constraints. 
\end{itemize}
    
\section{Related Work}
\label{sec:related_work}

\begin{figure*}[!htbp]
     \centering

    \begin{subfigure}[b]{0.45\linewidth}
      \centering
    \includegraphics[width=\textwidth]{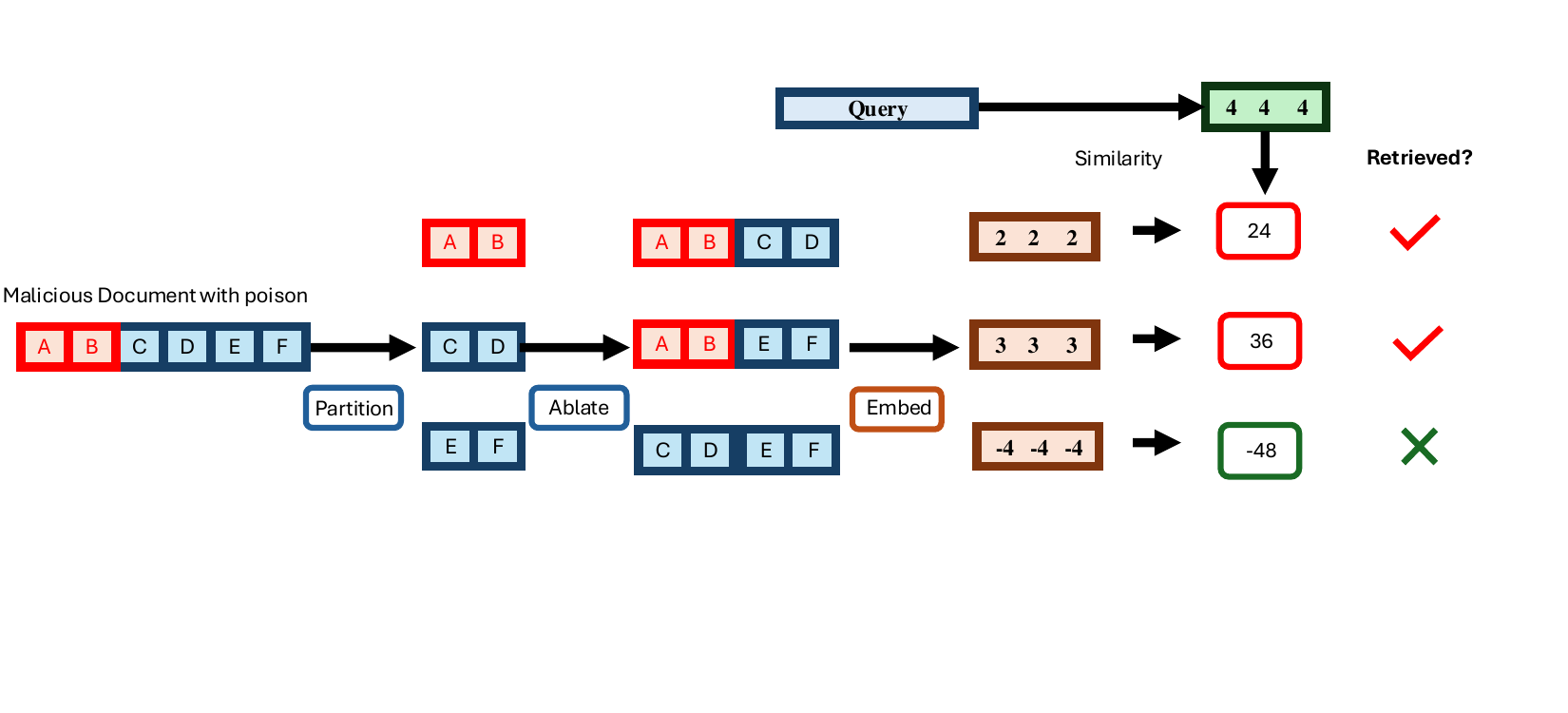} 
    \caption{\textbf{Naive Combination Baseline}}
    \end{subfigure}
    \hspace{10px}
    \begin{subfigure}[b]{0.45\linewidth}
    \centering
    \includegraphics[width=\textwidth]{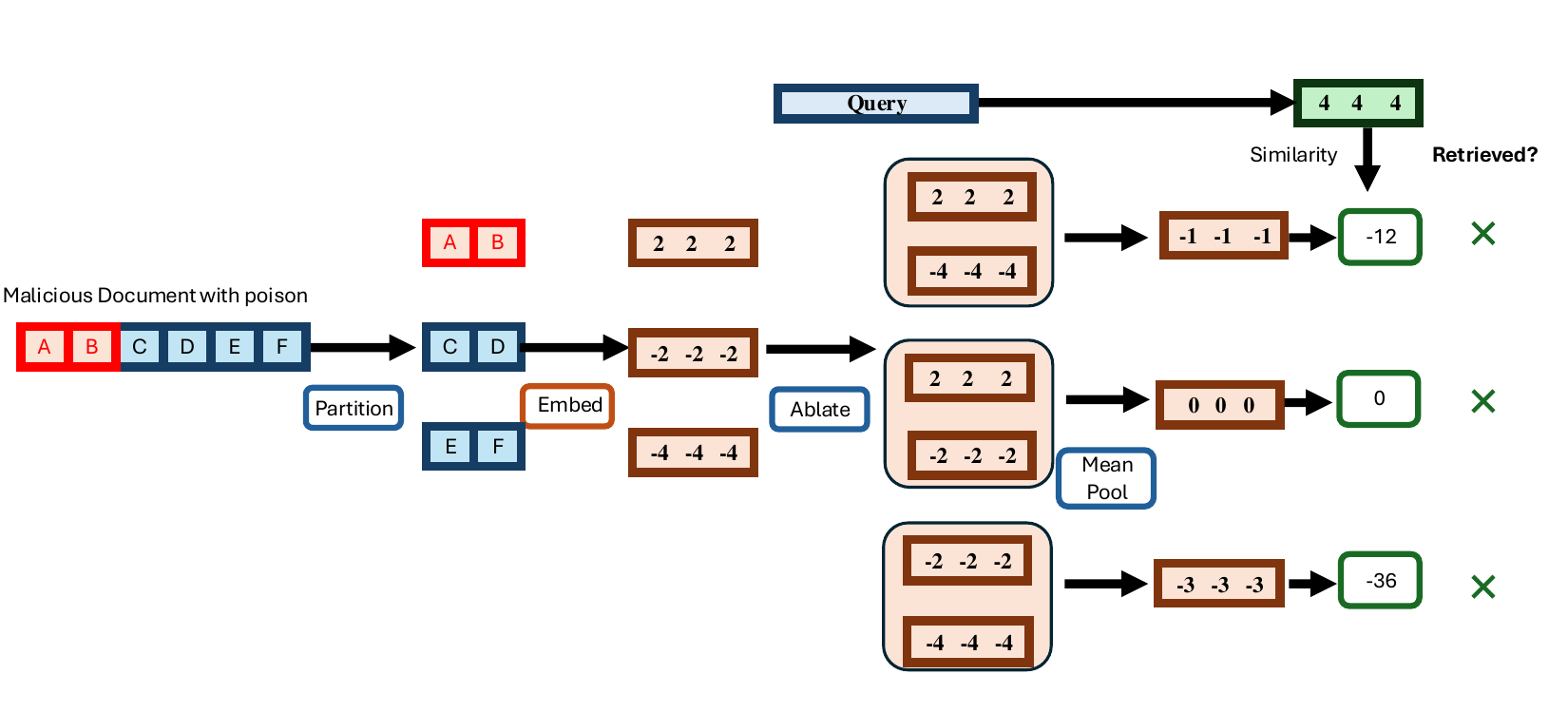}  
    \caption{\textbf{\defpart}}
    \end{subfigure}
    
    \caption{\textbf{Naive Combination Baseline vs. \defpart}: A toy example with N=3 and k=2. In this scenario, the naive combination approach is likely to retrieve the malicious document, since the poison persists in the raw text of many combined fragments (before embedding). In contrast, \defpart benefits from additional robustness due to mean pooling over fragment embeddings, which can dilute the effect of poisoned fragments and prevent the malicious document from being retrieved under the same conditions. For illustration purposes, we assume that large inner product values (e.g., 24, 36) correspond to the document ending up among the top-p retrieved results, while small values (e.g., –48, 0) correspond to cases where it does not.}  
     \label{fig:toy}
\end{figure*}

\textbf{Retrieval-Augmented Generation}: RAG improves the accuracy of LLM outputs and reduces hallucinations by retrieving relevant documents from a knowledge database given a query, and generating responses conditioned on these retrieved documents \citep{lewis2021retrievalaugmentedgenerationknowledgeintensivenlp, izacard2021leveragingpassageretrievalgenerative}. Particularly in sensitive fields such as law \citep{mao-etal-2024-rag}, RAG systems enable LLMs to generate reliable outputs that reflect up-to-date knowledge. This effectiveness has led to the wide deployment of RAG on both public and e-commerce platforms. Although traditional retrieval frameworks such as TF-IDF \citep{SALTON1988513} and BM25 \citep{10.1561/1500000019} have shown promise in retrieving relevant documents using word frequency statistics, recent advances in transformer-based dense retrievers \citep{izacard2021leveragingpassageretrievalgenerative, xiong2020approximatenearestneighbornegative, wang2024multilinguale5textembeddings, li2023generaltextembeddingsmultistage}—which encode semantic meanings into embedding vectors—have demonstrated superior performance on state-of-the-art retrieval benchmarks \citep{muennighoff2023mtebmassivetextembedding}.

\textbf{Attacks}: The corpus poisoning attacks against RAGs can be divided into black-box and white-box attacks based on the access the attacker has to the retriever model. The goal of the attacker is to either create a retrievable adversarial passage that can cause a harmful generation when added to the context of an LLM or craft poisons whose addition into the adversarial passage can make them retrievable for a given query. Works such as \citep{zou2024poisonedragknowledgecorruptionattacks, zhong2023poisoningretrievalcorporainjecting} have crafted white-box poisons by exploiting the gradient of the retrievers, which when added to an adversarial passage can fool the retriever into retrieving the passage. In black-box settings, the works of \citep{zou2024poisonedragknowledgecorruptionattacks} have proposed adding the query itself to the adversarial passage to make it retrievable.

\textbf{Defenses}: Early defenses, such as those by \cite{weller2024defendingdisinformationattacksopendomain}, proposed paraphrasing queries to retrieve multiple robust passages and thereby mitigate misinformation at the retrieval stage. Although these defenses can handle weaker adversaries, they often fail against stronger attacks \cite{zou2024poisonedragknowledgecorruptionattacks} and robust retrievers capable of preserving semantic meaning across paraphrases. Another line of work \citep{xiang2024certifiablyrobustragretrieval} proposes certified defenses against corpus poisoning at the generation stage (rather than at retrieval) by aggregating responses generated from each of the top-$p$ retrieved documents. However, these generation-stage defenses rely on strong assumptions—each golden document must independently suffice for generation, and retrievers must reliably retrieve an overwhelming number of golden documents. In practice, these conditions rarely hold, and such approaches are computationally expensive because long-context LLMs must be invoked multiple times per inference.

\section{Method}
\label{sec:method}
\subsection{Defense: \defpart}
Most state-of-the-art dense retrievers \cite{wang2024multilinguale5textembeddings, izacard2022unsuperviseddenseinformationretrieval, li2023generaltextembeddingsmultistage} are pre-trained using a large-scale contrastive loss \cite{khosla2021supervisedcontrastivelearning} and then fine-tuned on human-annotated data. Some retrievers, such as \cite{izacard2022unsuperviseddenseinformationretrieval}, select positive pairs during contrastive learning by sampling a random portion of a document and treating it as a positive example paired with the full document. This setup explicitly introduces an inductive bias that encourages the model to produce similar embeddings for different fragments of the same document. In practice, as we show in the Results section, even models that are not trained this way still exhibit similar behavior.

Inspired by this observation and deep partition-and-aggregation (DPA) defenses \cite{sun2022certifiablyrobustpolicylearning}, we propose \defpart. The key idea is to partition a document into $N$ fragments and apply the dense retriever's embedding model to each fragment individually. Due to the inductive bias of dense retrievers, the fragment embeddings tend to preserve the semantic similarity of the full document. We then average the embeddings of different combinations of fragments to form a final similarity score. If the number of poisoned fragments $n_p$ is not too large, their influence is diminished through the averaging step. By evaluating multiple such combinations and aggregating the results, \defpart effectively reduces the impact of the poisoned samples, as demonstrated in the Results section.

\begin{figure*}[!htbp]
    \centering
    \includegraphics[width=0.65\textwidth]{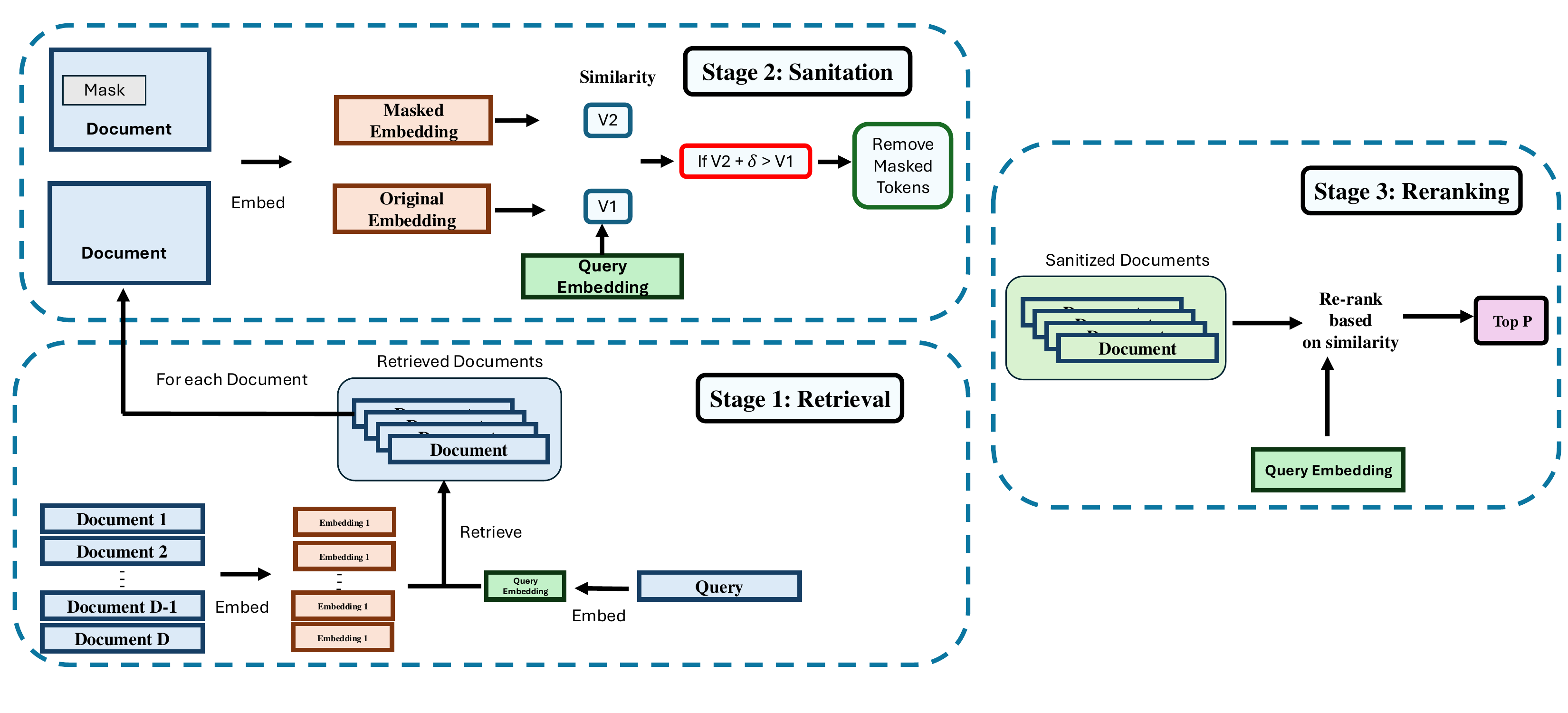} 
    \caption{\textbf{\defmask}: Figure illustrates the concept of \defmask, where the top $\alpha p$ retrieved documents are sanitized by observing the shift in similarity with the query under masking.}
    \label{fig:method_2}
\end{figure*}

In the context of dense retrieval as the first stage of RAG, and given the properties of document fragments, a document can be broken into $N$ fragments. Based on how we form combinations of these fragments, we consider two approaches, as seen below:

\begin{enumerate}
\item \textbf{\defpart}: We first embed each of the $N$ fragments using the dense retriever's embedding model. Then, we form combinations of $k$ fragments (e.g., all $\binom{N}{k}$ subsets of size $k$), and average their embeddings to form a combined representation. This approach leverages the inductive bias of dense retrievers, which produce similar embeddings for semantically related text, allowing the averaged embedding to preserve the original document's meaning even when some fragments may be poisoned. This method is illustrated in Figure \ref{fig:method_1}.
\item \textbf{Naive combination of fragments baseline}: In this approach, we first select combinations of $k$ out of $N$ document fragments (without first embedding them) and concatenate the raw text. The resulting text is then passed through the embedding model. A toy example containing a comparison between the two methods is shown later in Figure \ref{fig:toy}. Since embedding happens after mixing the content, poisoned fragments can---by design---have a significant influence on the final embedding.
\end{enumerate}

Once a new set of $\binom{N}{k}$ embeddings is formulated to represent each document, we can build $\binom{N}{k} \cdot D$ embedding databases for a given corpus of $D$ documents. These databases are used to retrieve $\binom{N}{k}$ sets of top-$p$ documents. The retrieved results can then be aggregated to form the final top-$p$ documents to be passed to the generator. We explore two aggregation strategies below:

\begin{enumerate}
\item \textbf{Intersection-based aggregation}: From the $\binom{N}{k}$ sets of top-$p$ documents, we select only those documents that appear in all of the sets. If no document appears across all sets, we randomly choose $p$ documents from the union of the $\binom{N}{k} \cdot p$ retrieved documents. While this approach can reduce the chance of retrieving adversarial documents, it often severely impacts the success rate (SR). As a result, we do not use this strategy in most of our experiments.
    
\item \textbf{Majority vote-based aggregation}: From the $\binom{N}{k}$ sets of top-$p$ documents, we select the $p$ documents that appear most frequently. It is important to note that the ideal $k$ condition for robustness in \citep{sun2022certifiablyrobustpolicylearning}, which assumes $n_p < \frac{N-1}{2}$, no longer guarantees robustness in this context. This is because the aggregation now involves selecting the top-$p$ documents by majority vote across combinations, rather than making a decision per combination. Thus, even if the adversarial document is not the top-ranked in most combinations, it may still be retrieved due to frequency. While setting $p = 1$ can restore the original robustness guarantee, it can greatly reduce utility. 

\end{enumerate}
Furthermore, the framework in \cite{sun2022certifiablyrobustpolicylearning} assumes that an adversary can influence the output if it is present in any single fragment within a combination. To be able to weaken this assumption, we considered the \defpart framework that minimizes the impact of poisoned fragments even when they are included in some combinations.

Motivated by the shortcomings in majority vote-based aggregation, we analyze the effectiveness of the \defpart framework compared to the naive combination of fragments in suppressing the effect of adversarial poisons when they are present in fragment combinations. Adversarial tokens are typically designed to increase the likelihood of retrieval when included in a document. This behavior poses a disadvantage for the naive combination approach—since the adversary remains in the raw text before embedding, its influence is preserved, and the resulting embedding can still lead to the document being retrieved. In contrast, in the \defpart approach, each fragment is embedded independently and then averaged, reducing the influence of individual poisoned fragments. Since the poisons are not optimized to dominate the mean of multiple embeddings, their effectiveness naturally decreases. Designing adaptive poisons to counteract this is difficult: it would require crafting embeddings with unusually large norms, which is hard to achieve in practice and can be mitigated by anomaly detection in embedding space. This distinction between the naive combination method and \defpart is illustrated with a toy example using $N=3$ and $k=2$ in Figure \ref{fig:toy}.

\subsection{Defense: \defmask}

The idea of \defmask takes inspiration from perturbation-based defenses. The idea behind perturbation-based defenses \cite{yang2021raprobustnessawareperturbationsdefending} is to analyze the behavior of a given sample in the presence of perturbations and make decisions accordingly. In the case of RAG, if the addition of a poison to a document makes it retrievable, then masking or occluding that token should cause a substantial drop in the similarity score between the document and the intended query. We leverage this insight to design the \defmask defense, as shown in Figure \ref{fig:method_2}, in the following way.

Given a corpus of documents $D$ and a query $q$, we first convert them into embeddings using the retriever model. We then retrieve the top $\alpha p$ documents, where $\alpha > 1$. Assume that the length of a single document is $l_i$ tokens. Each document has an initial similarity score $v_i^q$ with respect to the query. For each of the top $\alpha p$ documents, we divide the document into $\frac{l_i}{m}$ segments, where $m$ is a predefined hyperparameter representing the mask length. We mask the document at each of these $\frac{l_i}{m}$ positions and recompute the similarity score between the query and the masked version of the document, denoted by ${v_i^{q}}'$. If ${v_i^{q}}' + \delta > v_i^q$, we keep the masked tokens in the document; otherwise, we discard them. After $\frac{l_i}{m} \cdot \alpha p$ such operations, we obtain a new set of partially masked documents, which we refer to as sanitized documents.

Since retrievers are generally much smaller than LLMs, these operations can be parallelized efficiently to maintain acceptable time complexity. We then recompute the similarity between the sanitized documents and the query $q$, re-rank the $\alpha p$ documents, and finally select the top $p$ as the final set of retrieved documents.

\subsection{Attack: AdvRAGgen}

In addition to the attacks discussed in the Experiments section, and inspired by the work of AdvBDGen \citep{pathmanathan2024advbdgenadversariallyfortifiedpromptspecific}, we propose an interpretable attack against RAG retrieval, called \textbf{AdvRAGgen}. The idea behind this attack is to use a general-purpose causal language model that takes in a query and an adversarial document and generates a paraphrase of the adversarial document such that it is retrieved.

The generator is trained via Direct Preference Optimization (DPO) using three feedback signals:
\begin{enumerate}
\item The semantic similarity between the original adversarial document and its generated paraphrase, measured by a semantic embedding model. This ensures the paraphrase maintains the intended content, satisfying the generation condition.
\item The similarity between the query and the paraphrase in the target retriever's embedding space. This ensures the paraphrase is retrievable, satisfying the retrieval condition.
\item The negative ROUGE-L score between the query and the paraphrase. This discourages trivial attacks that involve simply inserting the query into the adversarial document.
\end{enumerate}

\begin{figure*}[!htbp]
    \centering   \includegraphics[width=0.7\textwidth]{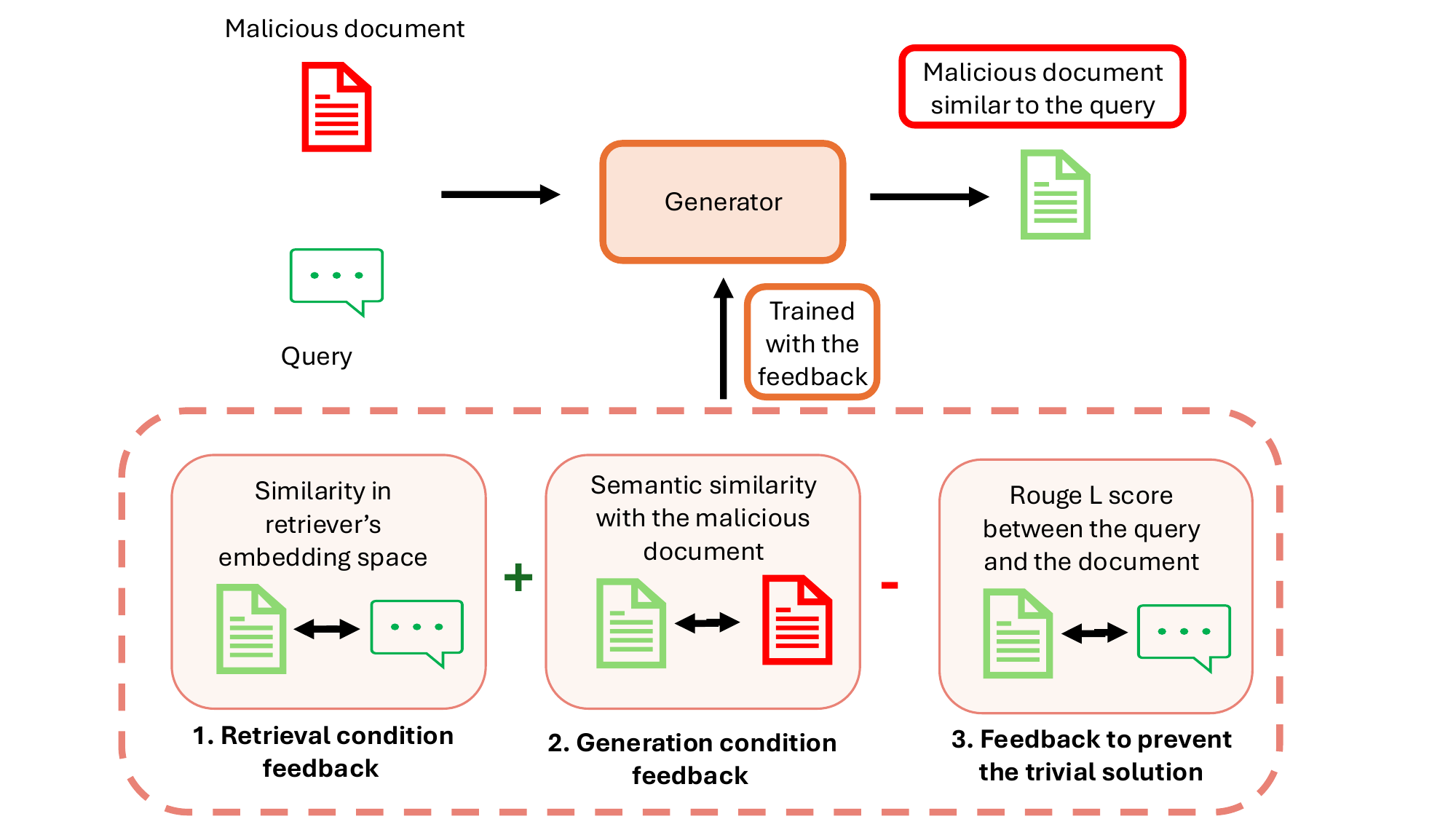} 
    \caption{\textbf{AdvBDGen style attack on retrieval}: This figure shows the overview of the poison generator's training framework inspired by AdvBDGen \citep{pathmanathan2024advbdgenadversariallyfortifiedpromptspecific}}
    \label{fig:advbdgen}
\end{figure*}

For further details, refer to the Appendix.
\section{Experiments}
\label{sec:exp}

\textbf{Dataset and Models}: We used two question answering datasets, namely Natural Question (NQ) \citep{kwiatkowski-etal-2019-natural} and Financial Opinion Mining and Question Answering (FIQA) \cite{10.1145/3184558.3192301}. NQ dataset is made of corpus of $2,681,468$ documents, while the FIQA is made  of a corpus of $57,638$ documents. From the queries, we randomly select $512$ queries and use them as our training set. For each query, we pick $3$ randomly pick documents and treat them as  irrelevant documents. The goal of the attacker is to craft poisons, whose addition will make these documents retrievable. Each of the queries has $10 <$ relevant queries, which we call golden documents in the corpus which is used to measure the utility of the defense under benign setting. As for dense retrievers we have considered $4$ retrievers namely, contriever \citep{izacard2022unsuperviseddenseinformationretrieval}, ANCE \cite{xiong2020approximatenearestneighbornegative}, multilingual e5 \citep{wang2024multilinguale5textembeddings} and GTE large \cite{li2023generaltextembeddingsmultistage}.

\textbf{Evaluation metrics}: We measure two evaluation metrics to measure both the robustness of the defense under attacks and the utility of the it under benign settings. The success of an attack is measured by the \textbf{\textit{attack success rate (ASR)}} which measures the number of times the retriever retrieved  atleast one poison or malicious document  for test queries. The utility, on the other hand, is measured by the \textbf{\textit{ success rate (SR)}} that counts the number of times the retriever retrieved atleast one golden document for test queries. We measure the average drop in the SR as measure of the drop in utility (lower the better) and the average drop in ASR as a measure of the robustness of a defense (higher the better).

\textbf{Attacks}: We consider both gradient-based and interpretable attacks as candidates for adversaries. Similar to the works of \citep{zou2024poisonedragknowledgecorruptionattacks, zhong2023poisoningretrievalcorporainjecting} as a candidate for the gradient based attack we consider the Hotfip \citep{ebrahimi2018hotflipwhiteboxadversarialexamples} style attacks which generate poison by searching for successful adversarial tokens in the token space guided by the gradient of the attacker's objective. We also consider a variant of this attack where the attacker instead of adding an adversarial token in a certain part of the text has the ability to spread out the tokens throughout the document which we call as HotFlip (spread out). As a candidate for interpretable attacks similar to \cite{zou2024poisonedragknowledgecorruptionattacks}, we add the query itself in the document as a poison which we call as query as poison. Furthermore we also propose a modfied version of AdvBDGen \cite{pathmanathan2024advbdgenadversariallyfortifiedpromptspecific} as mentioned in Section \ref{sec:method} as an additional interpretable attacks. 

\textbf{Baseline defenses}: Similar to the work of \citep{zou2024poisonedragknowledgecorruptionattacks}, we consider paraphrase and perplexity based defenses as the baseline. The idea behind paraphrase based defense is that certain poisons added to a document can be broken by paraphrasing the document before retrieval. We use LLama 3.3  70B \citep{grattafiori2024llama3herdmodels} as the paraphraser. Perplexity-based rely on the idea that the addition of poisons in a document can increase the perplexity of the document which can inturn be used to remove the adverserial document. We measure the perplexity here using a GPT2 model \citep{Radford2019LanguageMA}.
\section{ Results}
\label{sec:results}

\begin{table*}[!htbp]
  \caption{\textbf{Perplexity-based Defense}: This table shows that, similar to paraphrase-based defenses, perplexity-based defenses are effective at detecting gradient-based attacks. However, they fail to distinguish poisoned documents from benign ones in the case of interpretable attacks such as Query-as-Poison and AdvRAGgen, and therefore perform worse than the proposed defenses. We evaluate perplexity using four retrievers—Contriever \cite{izacard2022unsuperviseddenseinformationretrieval}, ANCE \cite{xiong2020approximatenearestneighbornegative}, Multilingual E5 \cite{wang2024multilinguale5textembeddings}, and GTE Large \cite{li2023generaltextembeddingsmultistage}. In the table, we report perplexity scores and highlight detections in red when the defense correctly identifies a poisoned document as malicious, and in green when it incorrectly classifies it as benign.}  \label{tab:perplexity_defense_main}
  \centering
  \begin{tabularx}{\textwidth}{@{}c|c|*{5}{Y}@{}}
    \toprule
    Dataset & Retriever & No Poison & HotFlip & HotFlip (spread out) & Query-as-Poison & AdvRAGgen \\
    \midrule

    \multirow{4}{*}{\makecell[l]{\textbf{Natural Questions (NQ)}\\ \footnotesize}}
      & \textbf{Contriever}      & \good{143} & \bad{989}  & \bad{1827} & \good{119} & \good{74} \\
      & \textbf{ANCE}           & \good{143} & \bad{5726} & \bad{12021} & \good{119} & \good{74} \\
      & \textbf{Multilingual E5}& \good{143} & \good{113} & \bad{392}  & \good{119} & \good{74} \\
      & \textbf{GTE Large}      & \good{143} & \bad{224}  & \bad{447}  & \good{119} & \good{74} \\
    \midrule

    \multirow{4}{*}{\makecell[l]{\textbf{FiQA}\\ \footnotesize}}
      & \textbf{Contriever}      & \good{143} & \bad{274}  & \bad{631}  & \good{100} & \good{53} \\
      & \textbf{ANCE}           & \good{143} & \bad{466}  & \bad{1095} & \good{100} & \good{53} \\
      & \textbf{Multilingual E5}& \good{143} & \good{86}  & \bad{252}  & \good{100} & \good{53} \\
      & \textbf{GTE Large}      & \good{143} & \good{113} & \bad{303}  & \good{100} & \good{53} \\
    \bottomrule
  \end{tabularx}
\end{table*}

Due to space constraints, we present both the ASR and SR (utility) results as averages over the four retrievers considered. For fine-grained results, refer to the Appendix. In the Appendix, we further provide hyperparameter analysis, motivating the choice of $N, k$ in \defpart and $\delta, m$ in \defmask. 

\begin{figure}[!htbp]
    \centering
    \begin{subfigure}[b]{0.43\linewidth}
        \centering    \includegraphics[width=\textwidth]{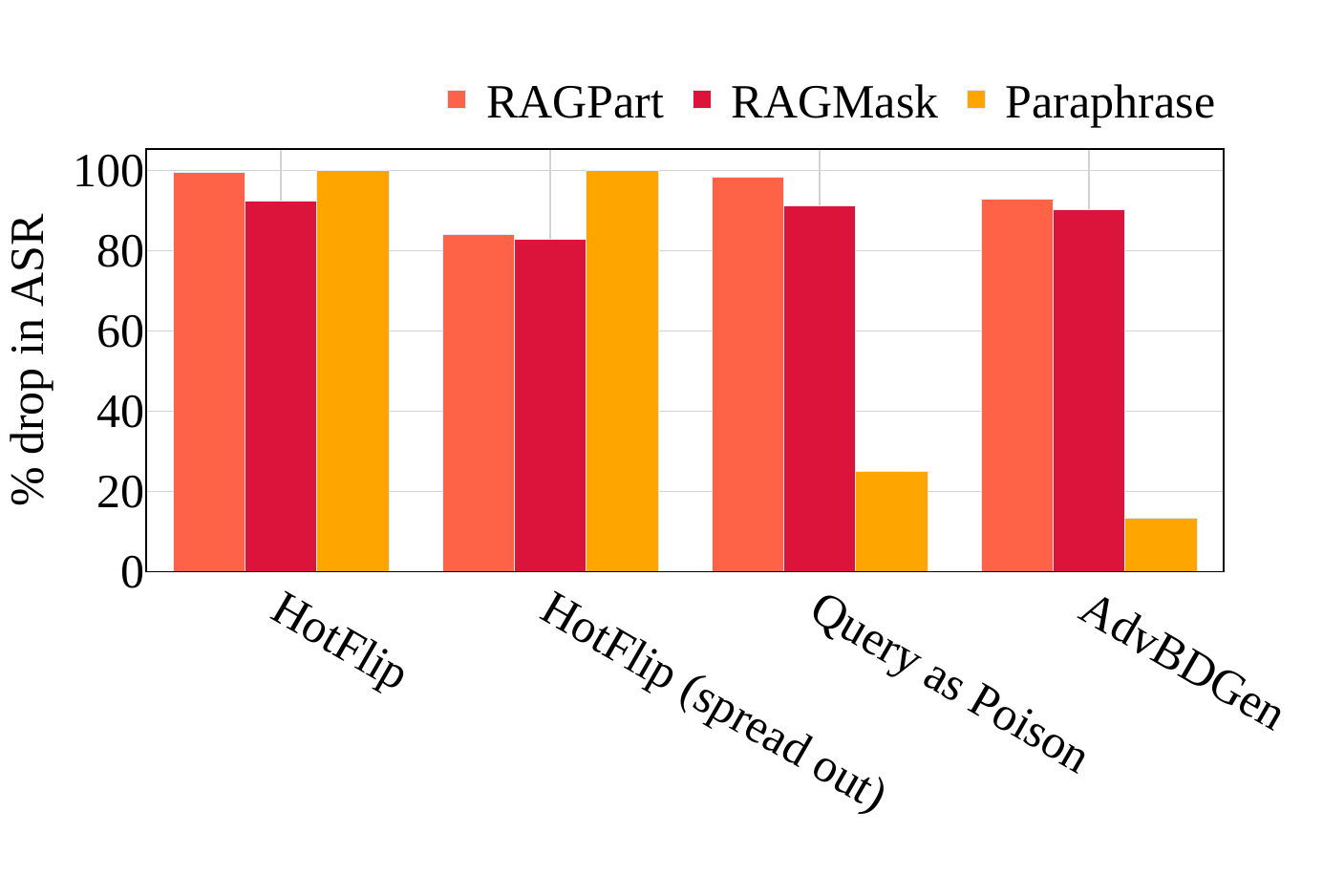} 
        \caption{\textbf{NQ dataset}}
    \end{subfigure}
    \hfill
    \begin{subfigure}[b]{0.43\linewidth}
        \centering
        \includegraphics[width=\textwidth]{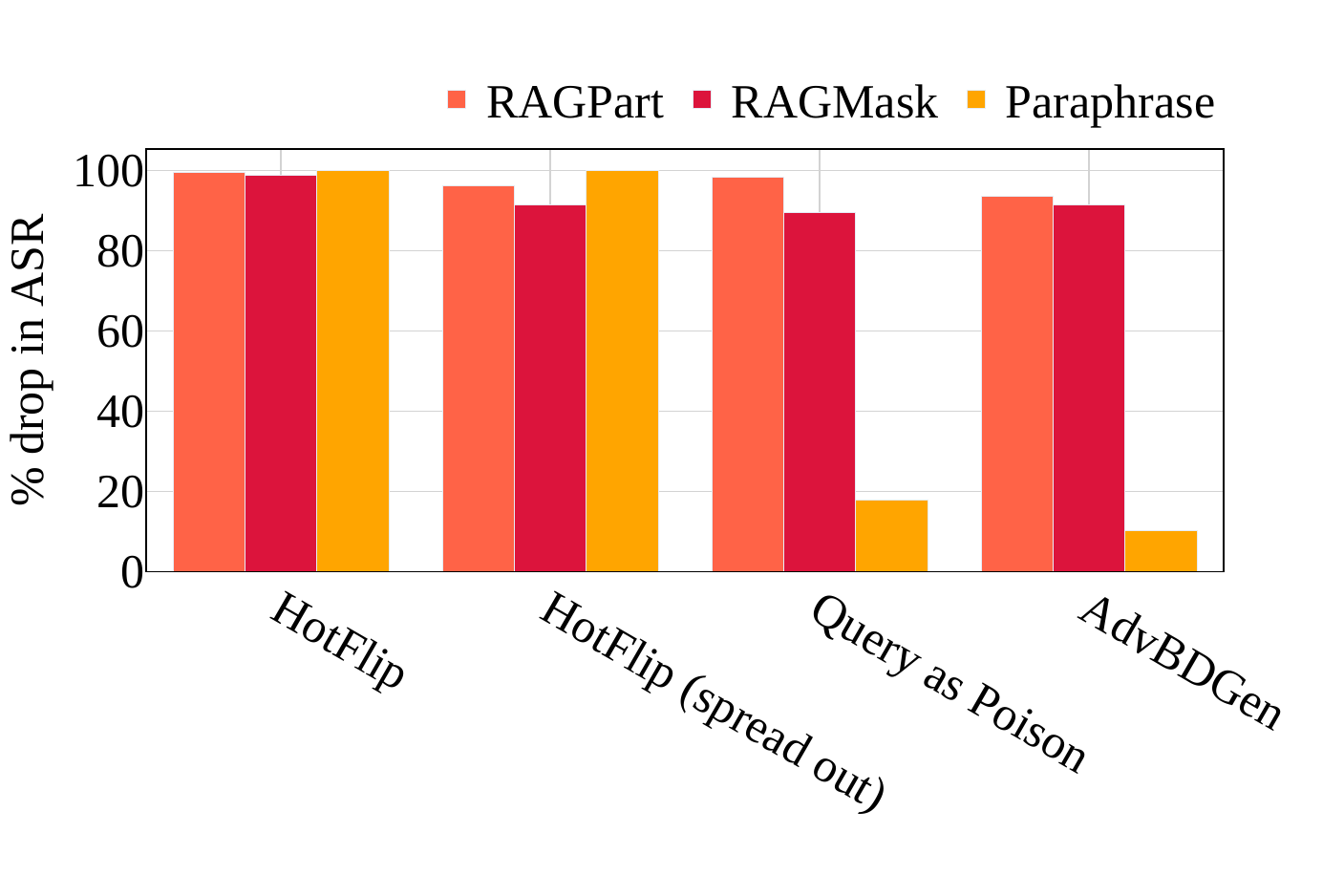} 
        \caption{\textbf{FiQA dataset}}
    \end{subfigure}
    \caption{\textbf{Drop in attack success rate (ASR) \textit{(higher is better)}}: This figure shows that across different datasets and attack types, both of our proposed methods—\defpart and \defmask—consistently maintain robustness. While paraphrasing-based defenses perform well against gradient-based attacks, they fail to defend against interpretable attacks such as Query-as-Poison and AdvRAGgen.}
    \label{fig:asr_main}
\end{figure}

\textbf{Ineffectiveness of the baseline defenses}: As seen in Figure \ref{fig:asr_main} and Table \ref{tab:perplexity_defense_main}, both the paraphrase-based and perplexity-based defenses fail to defend against interpretable poisoning attacks, even though they are effective against gradient-based attacks.

\begin{figure}[!htbp]
    \centering
    \begin{subfigure}[b]{0.43\linewidth}
        \centering
        \includegraphics[width=\textwidth]{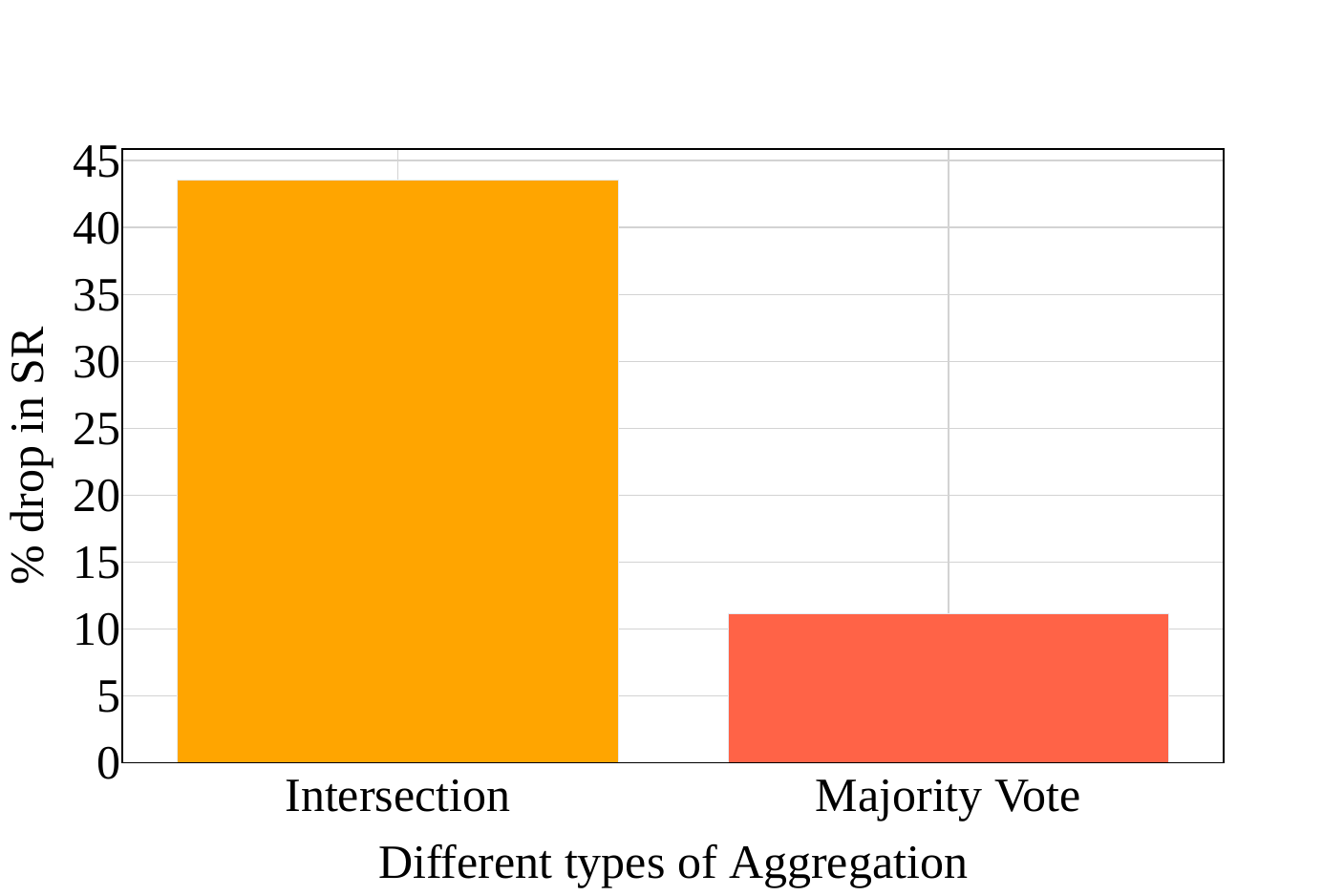} 
        \caption{\textbf{Naive}: Intersection-based aggregation vs majority vote-based aggregation}
    \end{subfigure}
    \hfill
    \begin{subfigure}[b]{0.43\linewidth}
        \centering
    \includegraphics[width=\textwidth]{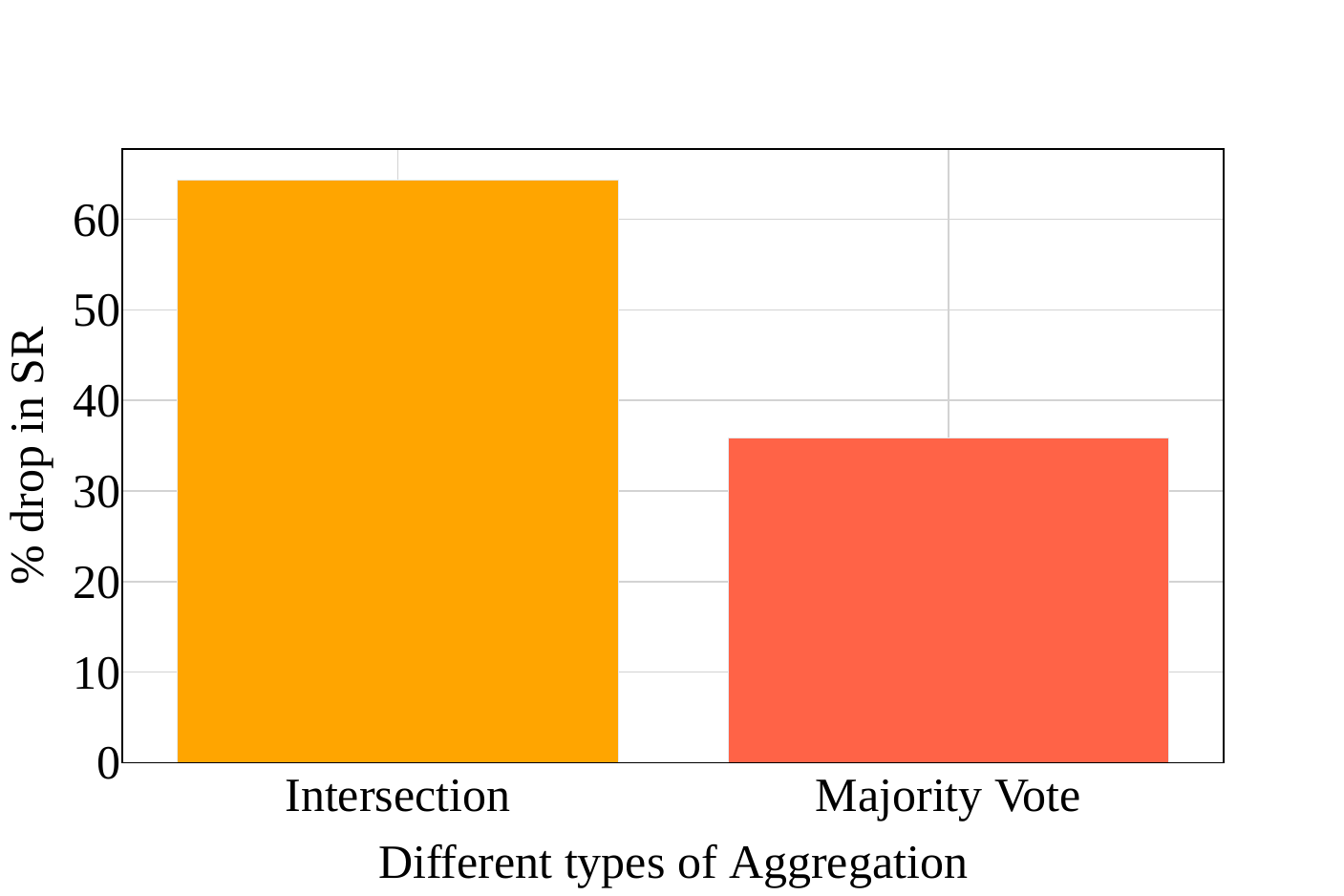} 
        \caption{\textbf{\defpart }: Intersection-based aggregation vs majority vote-based aggregation}
    \end{subfigure}

    \caption{\textbf{Drop in success rate (SR) on FiQA dataset \textit{(lower is better)} — Comparison of aggregation methods}: This figure shows that in both the naive combination of fragments and \defpart, intersection-based aggregation can lead to a larger drop in utility (SR), making it a less ideal choice for practical defenses.}
    \label{fig:sr_aggregation comparison}
\end{figure}

 \begin{figure}[!htbp]
    \centering
    \begin{subfigure}[b]{0.43\linewidth}
        \centering        \includegraphics[width=\textwidth]{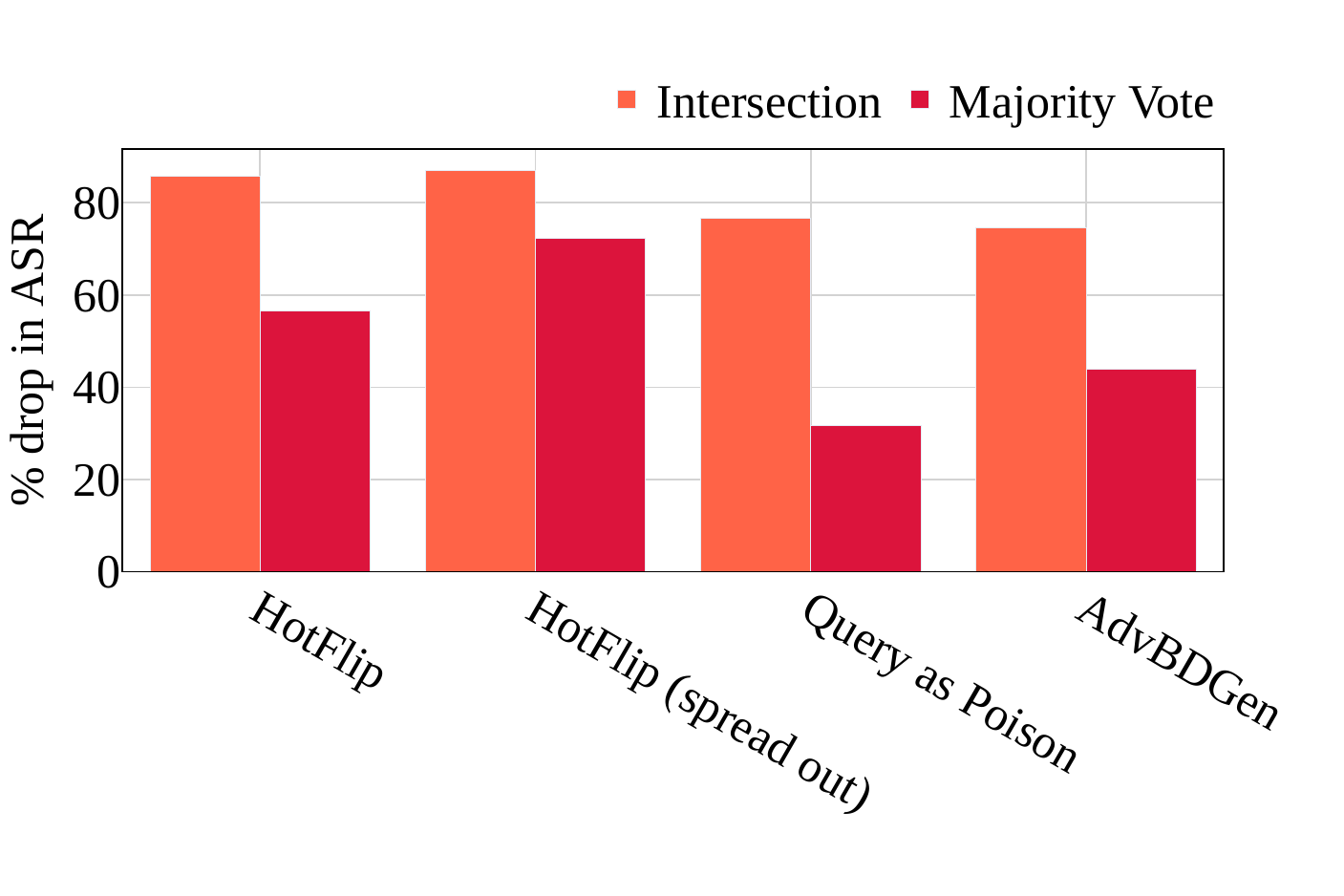} 
        \caption{\textbf{Naive combination}: Intersection-based aggregation vs majority vote-based aggregation}
    \end{subfigure}
        \hfill
    \begin{subfigure}[b]{0.43\linewidth}
        \centering        \includegraphics[width=\textwidth]{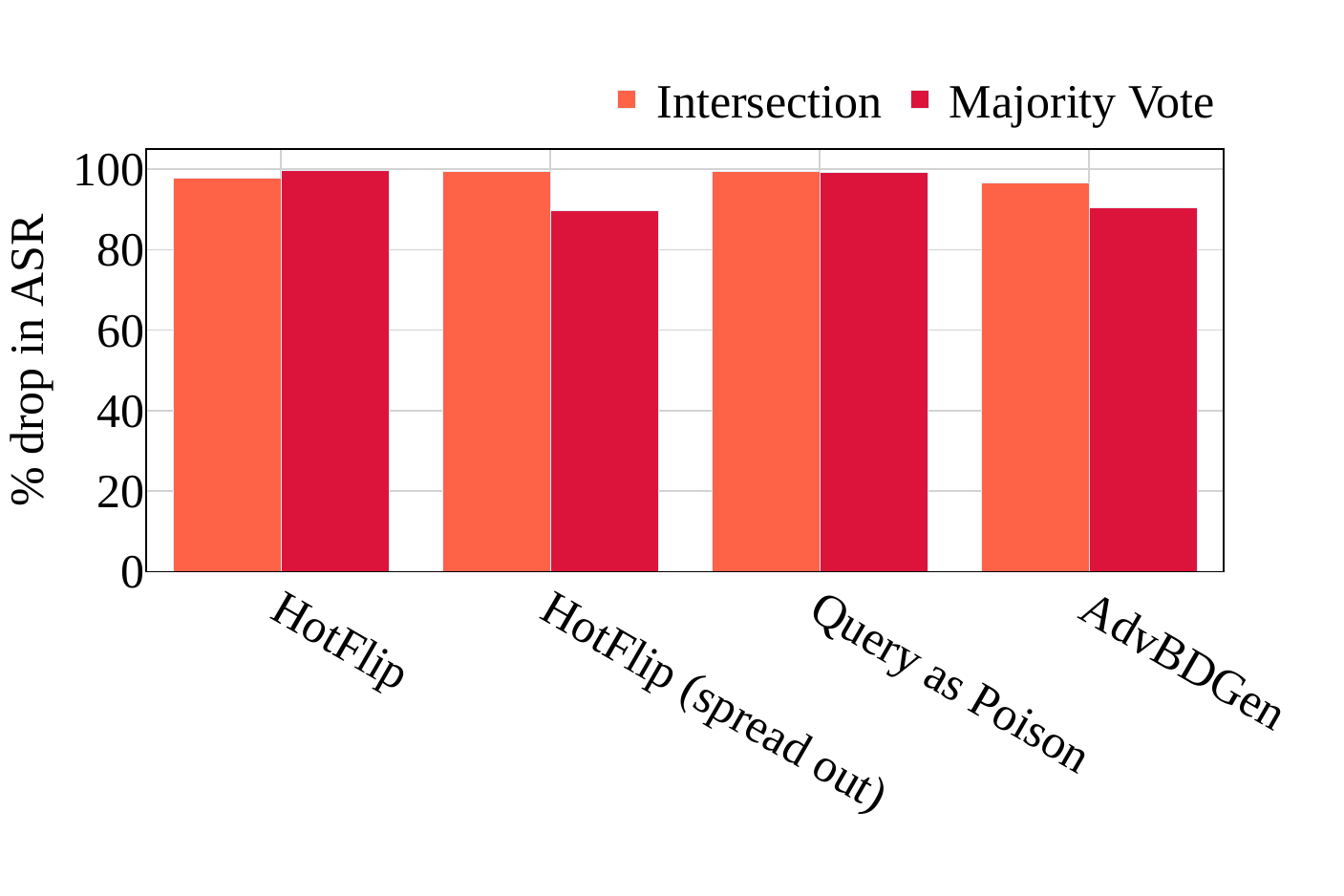} 
        \caption{\textbf{\defpart}: Intersection-based aggregation vs majority vote-based aggregation}
    \end{subfigure}

    \caption{\textbf{Drop in attack success rate (ASR) on FiQA dataset \textit{(higher is better)} — Comparison of aggregation methods}: This figure shows that for the naive combination of fragments, achieving acceptable ASR reduction often requires intersection-based aggregation, which comes at the cost of utility. In contrast, \defpart remains robust under both aggregation strategies, making it a more practical defense.}
    \label{fig:asr_aggregation comparison}
\end{figure}

\textbf{Effect of intersection-based aggregation and majority vote-based aggregation}: Intersection-based aggregation can remove poisoned documents more effectively than majority vote-based aggregation. This is shown in Figure \ref{fig:asr_aggregation comparison}, where intersection-based aggregation leads to a larger drop in ASR for both the naive combination baseline and \defpart. Due to the additional robustness of \defpart, the difference between intersection-based and majority vote-based aggregation is minimal in that case, unlike in the naive baseline. However, when considering SR over golden documents, Figure \ref{fig:sr_aggregation comparison} shows that intersection-based aggregation causes a larger drop in SR, making it unsuitable for practical deployment. Therefore, we adopt majority vote-based aggregation for the rest of the paper. For more detailed results, see Table \ref{tab:pre_sr_intersection_v_voting_pae}, \ref{tab:pre_asr_intersection_v_voting}, \ref{tab:post_sr_intersection_v_voting_}, and \ref{tab:post_asr_intersection_v_voting} in the Appendix.

\textbf{Viability of the naive combination baseline vs \defpart as a practical defense}: Although the naive combination baseline preserves utility slightly better than \defpart under majority vote-based aggregation (Figure \ref{fig:sr_aggregation comparison}), it fails to defend against any of the considered attacks. This makes \textit{\defpart a practically viable defense.}

\begin{figure}[!htbp]
    \centering
    \includegraphics[width=0.33\textwidth]{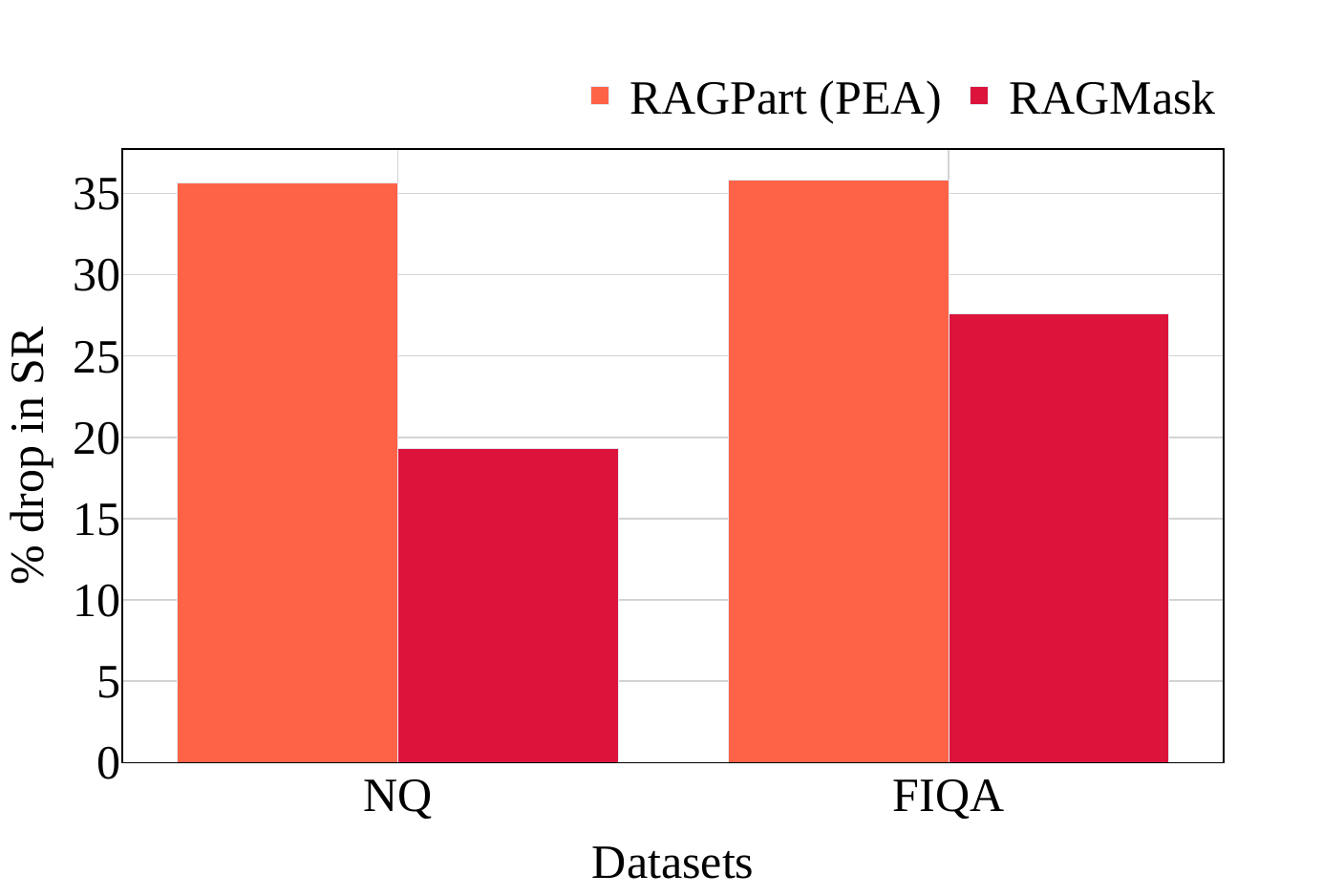} 
    \caption{\textbf{Drop in success rate (SR) \textit{(lower the better)}}: Figure showcases that both the \defpart and \defmask showed smaller drop in utility with \defmask performing slightly better. }
    \label{fig:sr_main}
\end{figure}

\textbf{Effectiveness of \defpart \& \defmask}: As shown in Figure \ref{fig:asr_main}, both \defpart and \defmask effectively defend against all considered attacks, outperforming baseline defenses, with \defpart achieving slightly better ASR reduction. In terms of utility preservation, \defmask demonstrates a stronger ability to maintain retrieval performance, as shown in Figure \ref{fig:sr_main}. 

\textbf{Computational efficiency of \defpart \& \defmask}: Although \defmask performs better in preserving the utility of benign retrievals and provides comparable robustness, it is relatively more expensive computationally. For a corpus of size $D$ and an embedding space of dimension $n_e$, \defpart requires $2n_e \cdot D \cdot \binom{N}{k}$ floating point operations (FLOPs), where $N$ is the number of fragments and $k$ is the combination size. Given that the retriever requires $R$ FLOPs to embed a document during the forward pass, \defmask requires $R \cdot \frac{l_i}{m} \cdot \alpha p$ FLOPs, where $l_i$ is the maximum document length, $m$ is the mask size, and $\alpha p$ is the number of documents retrieved prior to sanitation. In practice, $R$ may be large depending on the retriever's architecture. Thus, a computational tradeoff exists between \defpart and \defmask. However, since retrievers are generally smaller than autoregressive LLMs used for generation, \defmask remains computationally tractable compared to generation-stage defenses such as \cite{xiang2024certifiablyrobustragretrieval}, which require multiple LLM calls at inference time.

\section{Theoretical Analysis}
\label{sec:theory}

In the absence of adversarial documents, we assume that the top-$p$ retrieved results across the $\binom{N}{k}$ possible fragment mixes are consistent up to permutation. Let $n_a$ denote the number of adversarial documents in the corpus, each containing $n_p$ poisoned fragments. We assume $n_a \leq D - p$, ensuring that at least $p$ clean documents remain retrievable. We begin by analyzing the case where $n_a = 1$ (the extension to $n_a > 1$ is discussed later).
\\ \\
For each document, there are $\binom{N}{k}$ combinations of fragments that are either (1) fed into the embedding model after mixing (in the naive baseline), or (2) pre-embedded individually and averaged later (in \defpart{}).

When computing top-$p$ results (e.g., as shown in Figure~\ref{fig:method_1}), we construct a matrix of size $p \times \binom{N}{k}$, where each entry corresponds to a final embedding. In the absence of adversaries, each column would represent a mix of fragments from the same clean document (up to permutation).

However, in the presence of a single adversarial document, in the worst-case scenario, instead of observing $\binom{N}{k}$ occurrences for each of the top-$p$ clean documents, we may observe $\binom{N}{k}$ occurrences for $p - 1$ clean documents, $\binom{N}{k} - x$ occurrences for the $p$-th clean document, and $x$ occurrences for the poisoned mixes. For majority voting to be effective at filtering out poisoned embeddings, we require:
\[
x < \frac{1}{2} \binom{N}{k}.
\]

\paragraph{Naive Baseline.} In the naive combination-of-fragments baseline, embeddings are computed after fragment mixing. As a result, any mix that includes even a single poisoned fragment is considered adversarial. The number of such poisoned mixes is:
\[
x = \binom{N}{k} - \binom{N - n_p}{k}.
\]
This expression is derived by subtracting the number of fragment mixes that contain no poisoned fragment from the total number of possible mixes.

A sufficient condition for robustness in the naive baseline is:
\[
\binom{N}{k} - \binom{N - n_p}{k} < \frac{1}{2} \binom{N}{k}.
\]

\paragraph{\defpart{}.} In contrast, \defpart{} computes embeddings for individual fragments before mixing, and aggregates them via mean pooling. This reduces the influence of any single poisoned fragment. Suppose that a mix is considered poisoned only if it contains \emph{at least two} poisoned fragments. Then, the number of poisoned mixes is:
\[
x = \binom{N}{k} - \binom{N - n_p}{k} - n_p \binom{N - n_p}{k - 1}.
\]
Here:
\begin{itemize}
  \item $\binom{N - n_p}{k}$ counts the number of clean mixes (no poisoned fragments),
  \item $n_p \binom{N - n_p}{k - 1}$ counts the number of mixes containing exactly one poisoned fragment,
\end{itemize}
and their sum represents the number of mixes that are not adversarial for \defpart.

Thus, a sufficient condition for robustness in \defpart{} is:
\[
\binom{N}{k} - \binom{N - n_p}{k} - n_p \binom{N - n_p}{k - 1} < \frac{1}{2} \binom{N}{k}.
\]

In Tables~\ref{tab:naive_np2} and~\ref{tab:defpart_np2}, we show the values of \( N \) and \( k \) for which the sufficient condition for robustness holds under the naive baseline and \defpart{}, respectively, for \( n_p = 2 \). Similar results for \( n_p = 3 \) are presented in Tables~\ref{tab:naive_np3} and~\ref{tab:defpart_np3}. We observe that \defpart{} yields significantly more combinations of \( N \) and \( k \) that satisfy the robustness condition compared to the naive baseline. Overall, we recommend using lower values of \( N \) and higher values of \( k \) in practice to minimize performance degradation.

\begin{table}[H]
\centering
\caption{Sufficient condition for robustness under the naive baseline with \( n_p = 2 \). A green checkmark (\cmark) indicates that the condition holds for the given pair of \( N \) and \( k \).}\label{tab:naive_np2}
\begin{tabular}{c|ccccccccccccc}
\toprule
$N \backslash k$ & 3 & 4 & 5 & 6 & 7 & 8 & 9 & 10 & 11 & 12 & 13 & 14 & 15 \\
\midrule
3 & \xmark & N/A & N/A & N/A & N/A & N/A & N/A & N/A & N/A & N/A & N/A & N/A & N/A \\
4 & \xmark & \xmark & N/A & N/A & N/A & N/A & N/A & N/A & N/A & N/A & N/A & N/A & N/A \\
5 & \xmark & \xmark & \xmark & N/A & N/A & N/A & N/A & N/A & N/A & N/A & N/A & N/A & N/A \\
6 & \xmark & \xmark & \xmark & \xmark & N/A & N/A & N/A & N/A & N/A & N/A & N/A & N/A & N/A \\
7 & \xmark & \xmark & \xmark & \xmark & \xmark & N/A & N/A & N/A & N/A & N/A & N/A & N/A & N/A \\
8 & \xmark & \xmark & \xmark & \xmark & \xmark & \xmark & N/A & N/A & N/A & N/A & N/A & N/A & N/A \\
9 & \xmark & \xmark & \xmark & \xmark & \xmark & \xmark & \xmark & N/A & N/A & N/A & N/A & N/A & N/A \\
10 & \xmark & \xmark & \xmark & \xmark & \xmark & \xmark & \xmark & \xmark & N/A & N/A & N/A & N/A & N/A \\
11 & \cmark & \xmark & \xmark & \xmark & \xmark & \xmark & \xmark & \xmark & \xmark & N/A & N/A & N/A & N/A \\
12 & \cmark & \xmark & \xmark & \xmark & \xmark & \xmark & \xmark & \xmark & \xmark & \xmark & N/A & N/A & N/A \\
13 & \cmark & \xmark & \xmark & \xmark & \xmark & \xmark & \xmark & \xmark & \xmark & \xmark & \xmark & N/A & N/A \\
14 & \cmark & \xmark & \xmark & \xmark & \xmark & \xmark & \xmark & \xmark & \xmark & \xmark & \xmark & \xmark & N/A \\
15 & \cmark & \cmark & \xmark & \xmark & \xmark & \xmark & \xmark & \xmark & \xmark & \xmark & \xmark & \xmark & \xmark \\
\bottomrule
\end{tabular}
\end{table}

\begin{table}[H]
\centering
\caption{Sufficient condition for robustness under \defpart{} with \( n_p = 2 \). A green checkmark (\cmark) indicates that the condition holds for the given pair of \( N \) and \( k \).}\label{tab:defpart_np2}
\begin{tabular}{c|ccccccccccccc}
\toprule
$N \backslash k$ & 3 & 4 & 5 & 6 & 7 & 8 & 9 & 10 & 11 & 12 & 13 & 14 & 15 \\
\midrule
3 & \xmark & N/A & N/A & N/A & N/A & N/A & N/A & N/A & N/A & N/A & N/A & N/A & N/A \\
4 & \xmark & \xmark & N/A & N/A & N/A & N/A & N/A & N/A & N/A & N/A & N/A & N/A & N/A \\
5 & \cmark & \xmark & \xmark & N/A & N/A & N/A & N/A & N/A & N/A & N/A & N/A & N/A & N/A \\
6 & \cmark & \cmark & \xmark & \xmark & N/A & N/A & N/A & N/A & N/A & N/A & N/A & N/A & N/A \\
7 & \cmark & \cmark & \cmark & \xmark & \xmark & N/A & N/A & N/A & N/A & N/A & N/A & N/A & N/A \\
8 & \cmark & \cmark & \cmark & \xmark & \xmark & \xmark & N/A & N/A & N/A & N/A & N/A & N/A & N/A \\
9 & \cmark & \cmark & \cmark & \cmark & \xmark & \xmark & \xmark & N/A & N/A & N/A & N/A & N/A & N/A \\
10 & \cmark & \cmark & \cmark & \cmark & \cmark & \xmark & \xmark & \xmark & N/A & N/A & N/A & N/A & N/A \\
11 & \cmark & \cmark & \cmark & \cmark & \cmark & \xmark & \xmark & \xmark & \xmark & N/A & N/A & N/A & N/A \\
12 & \cmark & \cmark & \cmark & \cmark & \cmark & \cmark & \xmark & \xmark & \xmark & \xmark & N/A & N/A & N/A \\
13 & \cmark & \cmark & \cmark & \cmark & \cmark & \cmark & \cmark & \xmark & \xmark & \xmark & \xmark & N/A & N/A \\
14 & \cmark & \cmark & \cmark & \cmark & \cmark & \cmark & \cmark & \cmark & \xmark & \xmark & \xmark & \xmark & N/A \\
15 & \cmark & \cmark & \cmark & \cmark & \cmark & \cmark & \cmark & \cmark & \xmark & \xmark & \xmark & \xmark & \xmark \\
\bottomrule
\end{tabular}
\end{table}

\begin{table}[H]
\centering
\caption{Sufficient condition for robustness under the naive baseline with \( n_p = 3 \). A green checkmark (\cmark) indicates that the condition holds for the given pair of \( N \) and \( k \).}\label{tab:naive_np3}
\begin{tabular}{c|ccccccccccccc}
\toprule
$N \backslash k$ & 3 & 4 & 5 & 6 & 7 & 8 & 9 & 10 & 11 & 12 & 13 & 14 & 15 \\
\midrule
3 & \xmark & N/A & N/A & N/A & N/A & N/A & N/A & N/A & N/A & N/A & N/A & N/A & N/A \\
4 & \xmark & \xmark & N/A & N/A & N/A & N/A & N/A & N/A & N/A & N/A & N/A & N/A & N/A \\
5 & \xmark & \xmark & \xmark & N/A & N/A & N/A & N/A & N/A & N/A & N/A & N/A & N/A & N/A \\
6 & \xmark & \xmark & \xmark & \xmark & N/A & N/A & N/A & N/A & N/A & N/A & N/A & N/A & N/A \\
7 & \xmark & \xmark & \xmark & \xmark & \xmark & N/A & N/A & N/A & N/A & N/A & N/A & N/A & N/A \\
8 & \xmark & \xmark & \xmark & \xmark & \xmark & \xmark & N/A & N/A & N/A & N/A & N/A & N/A & N/A \\
9 & \xmark & \xmark & \xmark & \xmark & \xmark & \xmark & \xmark & N/A & N/A & N/A & N/A & N/A & N/A \\
10 & \xmark & \xmark & \xmark & \xmark & \xmark & \xmark & \xmark & \xmark & N/A & N/A & N/A & N/A & N/A \\
11 & \xmark & \xmark & \xmark & \xmark & \xmark & \xmark & \xmark & \xmark & \xmark & N/A & N/A & N/A & N/A \\
12 & \xmark & \xmark & \xmark & \xmark & \xmark & \xmark & \xmark & \xmark & \xmark & \xmark & N/A & N/A & N/A \\
13 & \xmark & \xmark & \xmark & \xmark & \xmark & \xmark & \xmark & \xmark & \xmark & \xmark & \xmark & N/A & N/A \\
14 & \xmark & \xmark & \xmark & \xmark & \xmark & \xmark & \xmark & \xmark & \xmark & \xmark & \xmark & \xmark & N/A \\
15 & \xmark & \xmark & \xmark & \xmark & \xmark & \xmark & \xmark & \xmark & \xmark & \xmark & \xmark & \xmark & \xmark \\
\bottomrule
\end{tabular}
\end{table}

\begin{table}[H]
\centering
\caption{Sufficient condition for robustness under \defpart{} with \( n_p = 3 \). A green checkmark (\cmark) indicates that the condition holds for the given pair of \( N \) and \( k \).}\label{tab:defpart_np3}
\begin{tabular}{c|ccccccccccccc}
\toprule
$N \backslash k$ & 3 & 4 & 5 & 6 & 7 & 8 & 9 & 10 & 11 & 12 & 13 & 14 & 15 \\
\midrule
3 & \xmark & N/A & N/A & N/A & N/A & N/A & N/A & N/A & N/A & N/A & N/A & N/A & N/A \\
4 & \xmark & \xmark & N/A & N/A & N/A & N/A & N/A & N/A & N/A & N/A & N/A & N/A & N/A \\
5 & \xmark & \xmark & \xmark & N/A & N/A & N/A & N/A & N/A & N/A & N/A & N/A & N/A & N/A \\
6 & \xmark & \xmark & \xmark & \xmark & N/A & N/A & N/A & N/A & N/A & N/A & N/A & N/A & N/A \\
7 & \cmark & \xmark & \xmark & \xmark & \xmark & N/A & N/A & N/A & N/A & N/A & N/A & N/A & N/A \\
8 & \cmark & \xmark & \xmark & \xmark & \xmark & \xmark & N/A & N/A & N/A & N/A & N/A & N/A & N/A \\
9 & \cmark & \cmark & \xmark & \xmark & \xmark & \xmark & \xmark & N/A & N/A & N/A & N/A & N/A & N/A \\
10 & \cmark & \cmark & \xmark & \xmark & \xmark & \xmark & \xmark & \xmark & N/A & N/A & N/A & N/A & N/A \\
11 & \cmark & \cmark & \cmark & \xmark & \xmark & \xmark & \xmark & \xmark & \xmark & N/A & N/A & N/A & N/A \\
12 & \cmark & \cmark & \cmark & \xmark & \xmark & \xmark & \xmark & \xmark & \xmark & \xmark & N/A & N/A & N/A \\
13 & \cmark & \cmark & \cmark & \cmark & \xmark & \xmark & \xmark & \xmark & \xmark & \xmark & \xmark & N/A & N/A \\
14 & \cmark & \cmark & \cmark & \cmark & \xmark & \xmark & \xmark & \xmark & \xmark & \xmark & \xmark & \xmark & N/A \\
15 & \cmark & \cmark & \cmark & \cmark & \cmark & \xmark & \xmark & \xmark & \xmark & \xmark & \xmark & \xmark & \xmark \\
\bottomrule
\end{tabular}
\end{table}

\subsection{\textbf{\defpart{} vs. Naive Baseline for Multiple Adversarial Documents}}

Assuming \( n_p \) is the number of poisoned fragments in the adversarial documents, we analyze the robustness of both methods under a worst-case scenario. Specifically, we assume that at most one poisoned mix embedding appears in each column of the \( p \times \binom{N}{k} \) matrix. While this assumption can be relaxed in practice, we adopt it here for analytical simplicity.

Under this setting, as long as the \emph{total number of poisoned mixes} across all \( n_a \) adversarial documents is less than \( \frac{n_a}{n_a + 1} \binom{N}{k} \), the least frequent clean document will still appear more often than any individual poisoned document in the final embedding matrix. This condition ensures robustness of top-\( p \) retrieval: each poisoned document can contribute at most \( \frac{1}{n_a + 1} \binom{N}{k} \) mixes, while the least frequent clean document will contribute more than \( \frac{1}{n_a + 1} \binom{N}{k} \). Thus, clean documents will dominate in frequency, and the top-\( p \) results will exclude all poisoned ones.

Accordingly, the sufficient conditions for robustness are:

\paragraph{Naive baseline:}
\[
\binom{N}{k} - \binom{N - n_p}{k} < \frac{1}{n_a + 1} \binom{N}{k}
\]

\paragraph{\defpart:}
\[
\binom{N}{k} - \binom{N - n_p}{k} - n_p \binom{N - n_p}{k - 1} < \frac{1}{n_a + 1} \binom{N}{k}
\]


\subsection{Computational Complexity: \defpart{} vs. Naive Baseline}

In this section, we analyze the computational complexity of computing the final document embeddings prior to similarity comparison with the query. We compare the naive baseline and \defpart{} from a theoretical perspective.

Assume that running the embedding model on a fraction \( \frac{1}{N} \) of a document requires \( R \) FLOPs. Then, running it on a fraction \( \frac{k}{N} \) (i.e., a mix of \( k \) fragments) requires \( k \times R \) FLOPs.

\paragraph{Naive Baseline.} For each document, there are \( \binom{N}{k} \) possible mixes of fragments. The embedding model is applied to each of these \( k \)-length mixes, so the total computational cost is:
\[
\text{FLOPs}_{\text{naive}} = D \times \binom{N}{k} \times (k \times R)
\]

\paragraph{\defpart{}.} In \defpart{}, the embedding model is applied once to each of the \( N \) individual fragments per document, for a total of:
\[
\text{FLOPs}_{\text{embedding}} = D \times N \times R
\]
Then, for each of the \( \binom{N}{k} \) combinations, we compute the mean of \( k \) precomputed embeddings, each of dimension \( n_e \), costing \( k \times n_e \) FLOPs per mix. Therefore, the total cost for the averaging step is:
\[
\text{FLOPs}_{\text{averaging}} = D \times \binom{N}{k} \times (k \times n_e)
\]
Summing both terms, the total FLOPs for \defpart{} is:
\[
\text{FLOPs}_{\defpart} = D \times N \times R + D \times \binom{N}{k} \times (k \times n_e)
\]

\paragraph{Example.} Suppose we have:
\[
R = 10^9 \quad (\text{FLOPs per embedding}), \quad D = 10^6, \quad N = 5, \quad k = 3, \quad n_e = 512
\]
Then:
\[
\binom{N}{k} = \binom{5}{3} = 10
\]

\begin{itemize}
  \item \textbf{Naive baseline:}
  \[
  \text{FLOPs}_{\text{naive}} = 10^6 \times 10 \times (3 \times 10^9) = 3 \times 10^{16}
  \]

  \item \textbf{\defpart{}:}
  \[
  \text{FLOPs}_{\defpart} = 10^6 \times 5 \times 10^9 + 10^6 \times 10 \times (3 \times 512) = 5 \times 10^{15} + 1.536 \times 10^{10}
  \]
\end{itemize}

Thus, \defpart{} reduces the dominant cost (embedding model inference) by a factor of \( \sim 6\times \), trading it for a much cheaper averaging step.

\section{Limitations}
\label{sec:limitation}

Despite showcasing robustness against attacks while maintaining a minimal drop in SR/utility, there are certain types of poison that the proposed methods cannot inherently defend against. For instance adversarial documents that are designed to be semantically similar to the query (e.g. for a query of "what is the capital of France?" the adversarial document of "The capital of France is Berlin") the proposed defenses cannot be used as a defense mechanism. These are poisons that we argue cannot be defended against in the retrieval stage because they can never be classified as poisons until a generation is made. Retrievers are inherently trained to encode only information about semantic similarity and not about the generation condition. For further discussion refer to Appendix \ref{A:QA}. 


\section{Conclusion}
\label{sec:conclusion}

In this work, we propose two retrieval-stage defenses, \defpart and \defmask, aimed at preventing adversarial documents from being retrieved in Retrieval-Augmented Generation (RAG) systems. Unlike generation-stage defenses, which often rely on strong assumptions—such as the self-sufficiency of each retrieved document, the presence of multiple relevant documents, high retriever accuracy, and the availability of significant computational resources—our defenses operate in a computationally tractable manner while maintaining robustness against a variety of poisoning attacks. Moreover, they preserve the utility of the retriever in benign settings.

We demonstrate that our methods outperform commonly used retrieval-based defenses, such as paraphrasing and perplexity filtering. Between our two approaches, \defmask offers better utility preservation and comparable robustness to attacks. However, we identify a tradeoff in computational cost, making \defpart a more practical choice in resource-constrained scenarios. 

Finally, while our defenses are effective against many corpus poisoning attacks, we also discuss in the appendix some attack types that remain difficult to detect at the retrieval stage, highlighting important directions for future research on developing defenses against such adversaries.

\section{Acknowledgments}
Pankayaraj, Panaitescu-Liess, and Huang are supported by DARPA Transfer from Imprecise and Abstract
Models to Autonomous Technologies (TIAMAT) 80321, National Science Foundation NSF-IIS-2147276 FAI, DODONR-Office of Naval Research under award number N00014-22-1-2335, DOD-AFOSR-Air Force Office of Scientific
Research under award number FA9550-23-1-0048, DOD-DARPA-Defense Advanced Research Projects Agency
Guaranteeing AI Robustness against Deception (GARD) HR00112020007, Adobe, Capital One and JP Morgan faculty
fellowships. Private support was provided by Peraton. 

\clearpage
\bibliography{main}
\newpage
\appendix
\addtocontents{toc}{\protect\setcounter{tocdepth}{2}}
\section{Q\&A}
\label{A:QA}
\subsection{Attack}

\begin{enumerate}
    \item \textbf{Why are the attacks success rates very low for gradient based methods (Hotflip and Hotflip (spread out) attacks?}

    Given a fixed number of adverserial tokens, gradient-based attacks that follow the paradigm of \citep{ebrahimi2018hotflipwhiteboxadversarialexamples} attacks work with two hyperparameters, namely the number of top candidates to consider based on the gradient and the number of iterations to the operations hotflip attack for. The higher these values are, the better the ASR is of these attacks. However, this also can lead to higher computations. Due to the number of experiments we were considering, we had limited these hyperparameters to 30 each. Stronger models such as multilingual e5 and GTE large needed more iterations to perform a successful attack. As an ablation, in Table \ref{tab:asr_abalation_grad}, we have provided results (for a smaller test size due to computation constraints, as creating one of these poison can take up-to an hour) for higher number iterations and proposed defense's efficacy against those poisons.   
\end{enumerate}

\subsection{Defense}

\begin{enumerate}
    \item \textbf{Generally, in deep-partition and aggregation (DPA) based defenses, don't people have a guarantee on robustness with majority voting-based aggregation? Why don't we see it in the case of RAGPart?}

    Those guarantees on majority voting are given in scenarios where only one decision is made from the aggregation. In contrast, in the case of RAG, generally the top $p$ samples are drawn from the set of documents. Even though the $N$ and $k$ are chosen in accordance with those guarantees, the adversary can end up getting chosen in the top $p$ documents even though it is not the topmost relevant document. While we can restrict ourselves to the top 1 document, it can severely hurt the performance of the system due to the existence of multiple documents and the retriever's deficiency.

    \item \textbf{What are the limitations of the proposed defenses? In what scenarios can the defense not defend and what is your argument against such scenarios?}

    The scenario we consider the most here is the scenario where there is an adversarial document and it is generally not retrievable for the retriever, and a poison is added to make it retrievable. If the adversarial document itself is retrievable, then our defenses may fail.
    
    For example, consider the following scenario. Given a query ``Where is Mount Everest located?'' a golden document can be ``Mount Everest is located between Nepal and Tibet,'' and an adversarial document can be ``Mount Everest is located at Spain.'' An important point to note here is that both these documents are semantically similar, i.e., both the documents talk about the location of Mount Everest, and one document is adversarial due to the fact that it contains misinformation or can induce the generation to contain misinformation. If the retriever is retrieving this document, that means the retriever doesn't have the knowledge about the factual error in the document. Thus, there is no way to defend against such a poison at the retrieval stage; rather, we argue that one should consider a generation-level defense as this is due to the deficiency of the retriever. If one is to solve this problem in a retrieval defense, then they should consider enhancing the retriever with more hard negatives (a document that is getting retrieved but is not actually a relevant document; i.e., harder to discriminate document). The attack we proposed (AdvRAGgen) to some level does create such a document (where it blends the query into the adversarial document), that's why it was a relatively harder attack to defend against, although our proposed defenses show an acceptable level of robustness against the attack.
    
    One may ask in what scenarios our assumption of the attack is practical. RAGs are used in e-commerce or in retrieving law documents in many cases. In the case of an e-commerce website, a malicious product seller may add a poison to make his product retrievable for an irrelevant query in order to increase their sales. Similarly, in the case of a legal RAG system, one may try to make a document with an incorrect judgment retrievable with malicious intentions.

\end{enumerate}

\newpage
\section{Attack}
\label{A:attack}
\subsection{AdvRAGgen}

We use an instruction-tuned Mistral 7B model \citep{jiang2023mistral7b} as the generator. Given a query and an irrelevant document from the training set, the generator is prompted to paraphrase the document into a retrievable form. Initially, the generator lacks knowledge of what is considered retrievable for a given query. To address this, we fine-tune the generator using three types of feedback (listed below), applying direct preference optimization (DPO) \citep{rafailov2024directpreferenceoptimizationlanguage}. Specifically, two paraphrases are generated and scored based on the feedback signals, and one is labeled as preferred. These preference pairs are then used to fine-tune the generator in an online DPO setup. The feedback signals are:

\begin{itemize}
    \item \textbf{Generation condition}: To ensure the generation condition is satisfied, the original malicious document must preserve its content when paraphrased. Therefore, we measure the semantic similarity between the original document and the generated response, and use this similarity score as a feedback signal for the generation objective.
    
    \item  \textbf{Retrieval condition}: For the malicious document to be successfully retrieved, it must be semantically similar to the query. To enforce this, we measure the similarity between the query and the document in the retriever’s embedding space and use this as a retrieval condition. While this constitutes a white-box attack by definition, we observe that the poisoned examples generated using one retriever model are transferable to others. In our experiments, we generate poisons using the Contriever model \cite{izacard2022unsuperviseddenseinformationretrieval} and find that they yield high attack success rates (ASR) even when applied to other retrievers.

    \item  \textbf{Preventing trivial solution}: In early experiments, optimizing with only the first two feedback signals caused the generator to copy the query into the document, effectively making the generated text identical to the query in the poison setting. To prevent this, we penalize such cases by measuring the Rouge-L score \citep{lin-2004-rouge} between the query and the generated document, applying a penalty when the score is high.
    
\end{itemize}

\begin{figure*}[!htbp]
    \centering
    \includegraphics[width=0.6\textwidth]{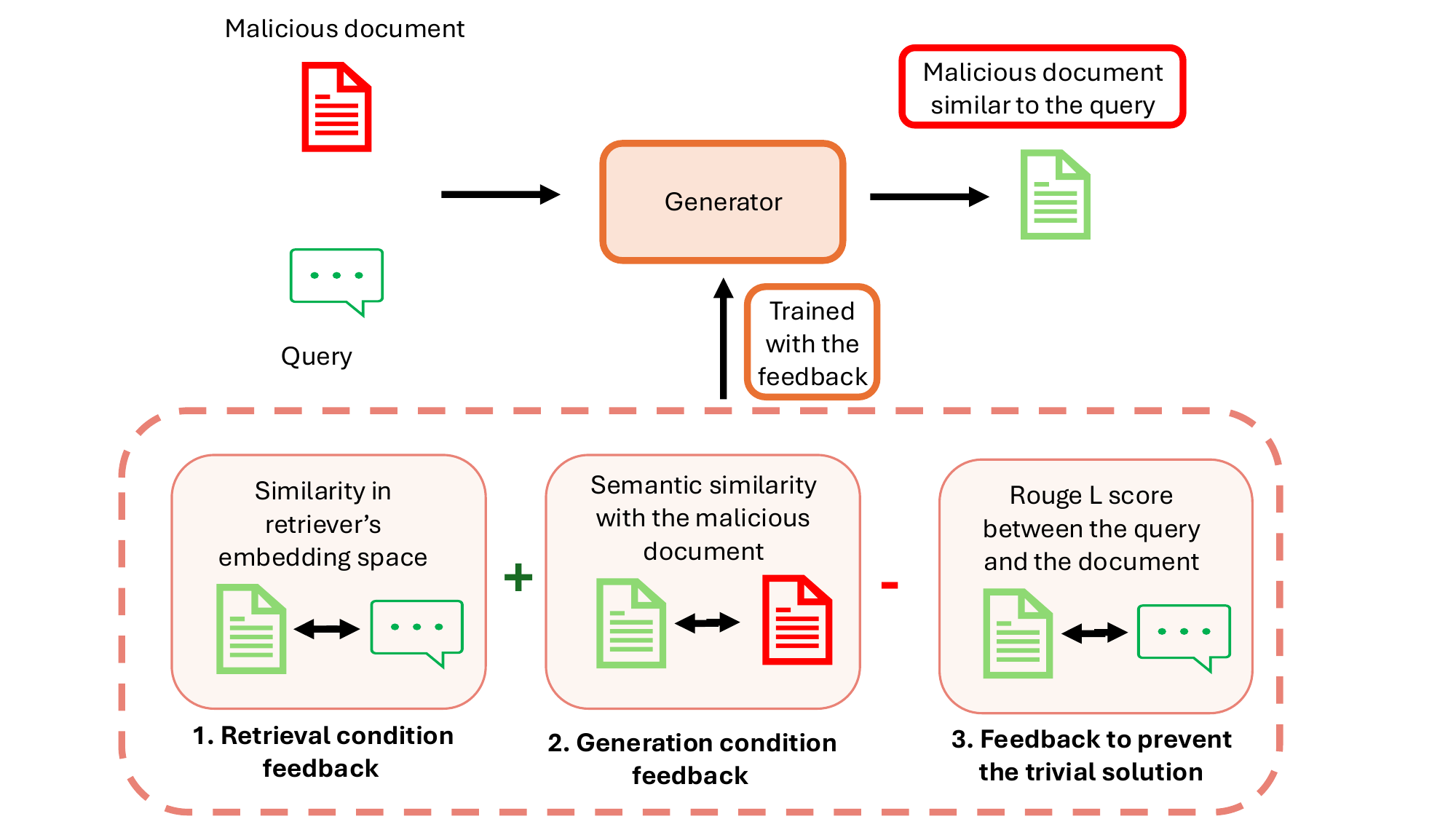} 
    \caption{\textbf{Pipeline of AdvRAGgen:} The figure illustrates the three feedback signals used to train AdvRAGgen. The \textit{generation condition} ensures similarity between the paraphrased and original malicious documents, the \textit{retrieval condition} enforces similarity between the query and the paraphrased document to enable successful retrieval, and the Rouge-L score serves as a \textit{regularizer} to prevent the trivial solution of copying the query as the poison.}
    \label{fig:advbdgen}
\end{figure*}

\newpage
\subsection{Ablation Results}

\begin{table}[!htbp]
  \caption{\textbf{ASR Hotflip with iterations (Ablation study)}: This table shows that for certain stronger models  such as multilingual e5 \citep{wang2024multilinguale5textembeddings} more iterations of the hotflip method is needed to produce an efficient poison and still the proposed defenses are capable of defending against the attack.}
  \label{tab:asr_abalation_grad}
  \centering
  \resizebox{\textwidth}{!}{\begin{tabular}{c|c|c|c|c|c|c|c|c|c}
    \toprule
    
    &  \multicolumn{8}{c}{\textbf{Natural Questions (NQ)} \cite{kwiatkowski-etal-2019-natural}}  \\
    \midrule
    
     & \multicolumn{8}{c}{ASR (\%)} \\
     \midrule
    Retriever & Defense  & itr=30 & itr=40  & itr=50   & itr=60  & itr=70  & itr=80  & itr=90  & itr=100  \\
    
    \midrule
     & No Defense & 16 & 50 & 50 & 78  & 82 & 88 & 88 & 90 \\ 
    & \defpart & 0 & 0 & 0 & 0 & 0 & 0 & 0 & 2 \\
   &  (N = 5, k=1) & & & & & & & & \\ 
   \textbf{Multilingual E5}  & \defpart   & 0 & 0 & 0 & 0 & 0 & 0 & 1 & 0 \\
   \cite{wang2024multilinguale5textembeddings}  &  (N = 5, k=3) & & & & & & & & \\ 
    & \defmask &2 & 2& 0& 2& 6 & 2 & 4 & 2 \\ 
   
    \bottomrule
  \end{tabular}}
\end{table}

\section{Defense}
\label{A:defense}

\subsection{Major Results}

\begin{table}[!htbp]
  \caption{\textbf{Success Rate}: This table shows the performance of \defpart and \defmask in benign retrieval scenarios. Here both the methods were able to preserve the utility on benign retrieval. }
  \label{tab:sr_main}
  \centering
  \resizebox{\textwidth}{!}{\begin{tabular}{c|cccc|cccc}
    \toprule
    
    &  \multicolumn{4}{c|}{\textbf{Natural Questions (NQ)} \cite{kwiatkowski-etal-2019-natural}} & \multicolumn{4}{c}{\textbf{FiQA} \cite{10.1145/3184558.3192301}}  \\
    \midrule
     & \multicolumn{4}{c|}{SR $\uparrow$ (\%)} & \multicolumn{4}{c}{SR  $\uparrow$(\%)}  \\
     \midrule
    Retriever  & No Defense & \defpart & \defpart  & \defmask & No Defense & \defpart & \defpart & \defmask \\
     & & (N = 5, k = 1 ) & (N = 5, k = 3 ) &  (m=1) &  &(N = 5, k = 1 )  &  (N = 5, k = 3) & (m=1) \\
     
    \midrule
    \textbf{Contriever} & \textbf{78} & 55 & 54 & 69 & \textbf{58} &  31 & 36 & 41\\
    \cite{izacard2022unsuperviseddenseinformationretrieval}& & & &  & & &\\
    \midrule
    \textbf{ANCE}  & \textbf{55} & 20 & 18 & 35 & \textbf{33} & 15 & 17 & 20 \\
    \cite{xiong2020approximatenearestneighbornegative} & & & & & &  & \\
    \midrule
    \textbf{Multilingual E5}  & \textbf{94} & 65 & 76 & 80 & \textbf{79} & 53 & 60 & 67 \\
    \cite{wang2024multilinguale5textembeddings} & & & & & & & \\ 
    \midrule
    \textbf{GTE Large}  & \textbf{69} & 32 & 49 & 59 & \textbf{64} & 28 & 43 & 47 \\
    \cite{li2023generaltextembeddingsmultistage} & & & & & & \\
    \bottomrule
  \end{tabular}}
\end{table}

\newpage
\begin{table}[!htbp]
  \caption{\textbf{Attack Success Rate}: This table shows the ability of \defpart and \defmask in defending against different types of attacks. Here both the methods were able to reduce the success rate across different types retrieval based attacks. }
  \label{tab:asr_main}
  \centering
    \resizebox{\textwidth}{!}{\begin{tabular}{c|c|cccc|cccc}
    \toprule
    
    \multicolumn{2}{c|}{} & \multicolumn{4}{c|}{Natural Questions (NQ) } & \multicolumn{4}{c}{FiQA }  \\
    \multicolumn{2}{c|}{} & \multicolumn{4}{c|}{ \cite{kwiatkowski-etal-2019-natural}} & \multicolumn{4}{c}{ \cite{10.1145/3184558.3192301}}  \\
    \midrule
     \multicolumn{2}{c|}{} & \multicolumn{4}{c|}{ASR $\downarrow$ (\%)} & \multicolumn{4}{c}{ASR $\downarrow(\%)$ }  \\
     \midrule
    Retriever & Attack Type & No Defense & \defpart &  \defpart & \defmask & No Defense & \defpart & \defpart & \defmask \\
    &  &  & (N=5, k=1)  &  (N=5, k=3)  & (m=10) &  & (N=5, k=1)  &  (N=5, k=3) & (m=10) \\

    \midrule
    &HotFlip &  87 & \textbf{2} & 0 & 7 & 95 & \textbf{2} & 0 & 1 \\
    \textbf{Contriever} & HotFlip (spread out)  & 85 & 5 & 4 & 9 & 92 & 2 & 6 & 3  \\
    \cite{izacard2022unsuperviseddenseinformationretrieval} &Query as poison & 78 & 2 & 3 & 4 & 85 & 9 & 1 & \textbf{3} \\
    &AdvRAGgen& 91 & 10 & 6 & 8 & 94 & 13 &11 & 7 \\
    \midrule

    &HotFlip & 61 & \textbf{0} & 1 & 4 & 78  & \textbf{0}  & 0 & 3 \\ 
    \textbf{ANCE} &HotFlip (spread out) & 45 & 1 & 0 & 6 & 67 & 6 & 1 & 3 \\ \cite{xiong2020approximatenearestneighbornegative}&Query as poison & 68 & 0 & 0 & 8 & 65 & 5 & 0 & \textbf{3} \\
    &AdvRAGgen & 79 & 6 & 3 & 8 & 86 & 6 &4 & 5 \\
    \midrule

    &HotFlip & 18 & \textbf{0} & 0 &1  & 21 & \textbf{0} & 1 & 0 \\
    \textbf{Multilingual E5} &HotFlip (spread out) & 15 & 0 & 8 & 3 & 26 & 0& 0 & 7\\
    \cite{wang2024multilinguale5textembeddings}&Query as poison & 73 & 3 & 0 & 9 & 45 & \textbf{9} & 0 & 7 \\
    &AdvRAGgen & 81 & 8 & 7 & 4 & 95 & 15 &14 & 7\\
    \midrule

    &HotFlip& 10 & \textbf{0}  & 0 & 1 & 20 & \textbf{0} & 0 & 0 \\ 
    \textbf{GTE Large} &HotFlip (spread out) & 16 & 1 & 1 & 4 & 28 & 0 & 2 & 0\\
    \cite{li2023generaltextembeddingsmultistage}& Query as poison & 63 & 2 & 2 & 4 & 58 & 5  & 1 &  \textbf{0} \\
    &AdvRAGgen & 70 & 6 & 5 & 7 & 83 & 8 &6 & 5 \\

    \bottomrule
  \end{tabular}}
\end{table}

\subsection{Hyperparameter Analysis: Effect of $N$, $k$ in \defpart}

\begin{table}[!htbp]
  \caption{\textbf{Hyperparameter analysis on \defpart (SR $\%$)}: In this table we show effect of $N$, $k$ in the \defpart in utility preservation. This ablation showcases the drawbacks of using larger $N$ which can lead to degradation the of the utility thus highlighting the importance of \defpart as opposed to naive aggregation. The original SR is \textbf{74$\%$}. Experiments were done on the FiQA dataset.}
  \label{tab:sr_part_hyperparameter}
  \centering
  \begin{tabular}{c|cccccc}
    \toprule
    & $k$ =1	& $k$ = 3	& $k$ = 5 & $k$ = 10 & 	$k$ =15 & $k$ = 20\\
    \midrule
    $N$ = 5 & 	53	& 60 & 	60 & 	N/A & 	N/A	& N/A \\
    $N$ = 10 &	44	& 45 & 	41 & 	39 & 	N/A	& N/A \\
    $N$ = 15 &	34	& 33 & 	30 & 	28 & 	28	& N/A \\
    $N$ = 20 & 	26	& 26 & 24 & 	24 & 	22	& 21 \\
    \bottomrule
  \end{tabular}
\end{table}

\begin{table}[!htbp]
  \caption{\textbf{Hyperparameter analysis on \defpart (ASR $\%$)}: In this table we show effect of $N$, $k$ in the \defpart in defense. Using higher k can lead to better defense against the attacks. Here the original attack success rate was at \textbf{95$\%$} under an AdvRAGgen semantic attack. This signifies the capability of the defense. Furthermore, \defpart was able to afford a larger $k$ which can result in lower computational overhead. Experiments were done on the FiQA dataset. }
  \label{tab:asr_part_hyperparameter}
  \centering
  \begin{tabular}{c|cccccc}
    \toprule
    & $k$ =1	& $k$ = 3	& $k$ = 5 & $k$ = 10 & 	$k$ =15 & $k$ = 20\\
    \midrule
    $N$ = 5 & 	15	& 14 & 	7 & 	N/A & 	N/A	& N/A \\
    $N$ = 10 &	19	& 14 & 	5 & 	1 & 	N/A	& N/A \\
    $N$ = 15 &	15	& 15 & 	8 & 	1 & 	1	& N/A \\
    $N$ = 20 & 	20	& 9 & 4 & 	1 & 	1	& 1 \\
    \bottomrule
  \end{tabular}
\end{table}

\newpage

\subsection{Hyperparameter Analysis: Effect of $m$, $\delta$ in \defmask}
\begin{table}[!htbp]
  \caption{\textbf{Hyperparameter analysis on \defmask (SR $\%$)}: In this table we show effect of $\delta$, $m$ in the \defmask in utility preservation. The original SR is \textbf{74$\%$}. Experiments were done on the FiQA dataset. Even though at larger $\delta$ \defmask was able to preserve utility as seen below it lead to a lesser effectiveness against stronger attacks such the proposed AdvRAGgen.}
  \label{tab:sr_mask_hyperparameter}
  \centering
  \begin{tabular}{c|cccc}
    \toprule
    &  $m$ = 10	& $m$ = 15 & $m$ = 20 & 	$m$ = 25 \\
    \midrule
    $\delta$ = 0.01 & 	66 & 65 & 	66 & 66 \\
    $\delta$ = 0.05 &	73 & 73 & 	72 & 71	\\
    $\delta$ = 0.1 &	74 & 	74 & 	74 & 	74	\\
    $\delta$ = 0.5 & 	74 & 74 & 	74 & 	74 \\
    \bottomrule
  \end{tabular}
\end{table}

\begin{table}[!htbp]
  \caption{\textbf{Hyperparameter analysis on \defmask (ASR $\%$)}: In this table we show effect of $\delta$, $m$ in the \defmask in defense. Here the original attack success rate was at \textbf{95$\%$} under an AdvRAGgen semantic attack. Larger mask sizes $m$ were shown to be ideal under smaller $\delta$. Experiments were done on the FiQA dataset.}
  \label{tab:asr_mask_hyperparameter}
  \centering
  \begin{tabular}{c|ccccc}
    \toprule
    & $m$ = 5	& $m$ = 10	& $m$ = 15 & $m$ = 20 & 	$m$ = 25 \\
    \midrule
    $\delta$ = 0.01 & 	11	& 7 & 	5 & 	6 & 	6 \\
    $\delta$ = 0.05 &	44	& 28 & 	18 & 	12 & 	9	\\
    $\delta$ = 0.1 &	55	& 49 & 	35 & 	27 & 	22	\\
    $\delta$ = 0.5 & 	57	& 57 & 57 & 	57 & 	57 \\
    \bottomrule
  \end{tabular}
\end{table}

\newpage
\subsection{Hyperparameter Analysis: Effect of \defpart vs Naive combination of fragments}

\begin{table}[!htbp]
  \caption{\textbf{RAGPart SR - \defpart vs Naive combination of fragments} in FiQA: The table shows that under majority voting aggregation, \defpart better preserves utility compared to a naive combination across multiple retrievers.}
  \label{tab:sr_pre_embed_v_post_embed}
  \centering
  \resizebox{\textwidth}{!}{\begin{tabular}{c|ccc}
    \toprule
    
    &  \multicolumn{3}{c}{\textbf{FiQA} \cite{10.1145/3184558.3192301}}  \\
    \midrule
     & \multicolumn{3}{c}{SR $\uparrow$ (\%)} \\
     \midrule
    Retriever  & No Defense & Naive combination of fragments  (N=5, k=3) & \defpart   (N=5, k=3) \\

    \midrule
    \textbf{Contriever} & \textbf{58} & 35 & 36 \\
    \cite{izacard2022unsuperviseddenseinformationretrieval}& & & \\
    \midrule
    \textbf{ANCE}  & \textbf{33} & 17 & 18\\
    \cite{xiong2020approximatenearestneighbornegative} & & & \\
    \midrule
    \textbf{Multilingual E5}  & \textbf{79} & 48 & 60 \\
    \cite{wang2024multilinguale5textembeddings} & & & \\ 
    \midrule
    \textbf{GTE Large}  & \textbf{64} &34 & 43 \\
    \cite{li2023generaltextembeddingsmultistage} & & & \\
    \bottomrule
  \end{tabular}}
\end{table}

\subsection{Effect of Aggregation methods in Naive combination of fragments}

\begin{table}[!htbp]
  \caption{\textbf{Naive combination of fragments SR - Intersection based aggregation vs majority vote based aggregation} in  FiQA: While intersection-based aggregation offers better efficiency against attacks, it also leads to a significant drop in utility, making it a less ideal choice for aggregation. }
  \label{tab:pre_sr_intersection_v_voting_pae}
  \centering
  \resizebox{\textwidth}{!}{\begin{tabular}{c|ccc}
    \toprule
    &  \multicolumn{3}{c}{\textbf{FiQA} \cite{10.1145/3184558.3192301}}  \\
    \midrule
     & \multicolumn{3}{c}{SR $\uparrow$ (\%)} \\
     \midrule
     Retriever & No Defense & Naive & Naive  \\
      & & combination  &  combination \\
     & & (N=5, k=3)  & (N=5, k=3) \\
     & & \textbf{Intersection based aggregation} & \textbf{Majority vote based aggregation } \\
    \midrule
    \textbf{Contriever} & \textbf{58} & 35 & 52 \\
    \cite{izacard2022unsuperviseddenseinformationretrieval}& & & \\
    \midrule
    \textbf{ANCE}  & \textbf{33} & 17 & 27\\
    \cite{xiong2020approximatenearestneighbornegative} & & & \\
    \midrule
    \textbf{Multilingual E5}  & \textbf{79} & 48 & 74 \\
    \cite{wang2024multilinguale5textembeddings} & & & \\ 
    \midrule
    \textbf{GTE Large}  & \textbf{64} & 34 & 52 \\
    \cite{li2023generaltextembeddingsmultistage} & & & \\
    \bottomrule
  \end{tabular}}
\end{table}

\begin{table}[!htbp]
  \caption{\textbf{ Naive combination of fragments ASR - Intersection based aggregation vs voting based aggregation} in FiQA:  Even though majority voting based aggregation results in better utility preservation under naive combination due to it's lack of additional robustness as in \defpart it ends up being ineffective as a defense thus making the naive combination of fragments as an impractical version of defense against retrival poisoning.}
  \label{tab:pre_asr_intersection_v_voting}
  \centering
  \resizebox{\textwidth}{!}{\begin{tabular}{c|c|ccc}
    \toprule
    & & \multicolumn{3}{c}{\textbf{FiQA} \cite{10.1145/3184558.3192301}}  \\
    \midrule
     & & \multicolumn{3}{c}{ASR $\downarrow$ (\%)} \\
     \midrule

      Retriever & Attack & No Defense & Naive & Naive  \\
    & & & combination & combination \\
    
     & & & N=5, k=3)  & N=5, k=3) \\
      & & & \textbf{Intersection based aggregation} & \textbf{Voting based aggregation } \\

    \midrule
    & HotFlip & \textbf{95} & 23  & 76 \\
    \textbf{Contriever} & HotFlip (spread out) & \textbf{92} & 21  & 47\\
    \cite{izacard2022unsuperviseddenseinformationretrieval}& Query as poison& \textbf{85} & 20 & 67\\
    & AdvRAGgen & \textbf{94} & 24 & 77 \\
    \midrule

    & HotFlip & \textbf{78} & 18 & 50 \\
    \textbf{ANCE}  & HotFlip (spread out) & \textbf{67} &  7 & 18\\
    \cite{xiong2020approximatenearestneighbornegative} & Query as poison & \textbf{65} & 15  & 42\\
   & AdvRAGgen & \textbf{89} & 20  & 58 \\
    \midrule

    & HotFlip & \textbf{21} & 2 & 3\\
    \textbf{Multilingual E5}  & HotFlip (spread out) & \textbf{26} & 3 & 4\\
    \cite{wang2024multilinguale5textembeddings} & Query as poison & \textbf{45} & 14 & 36 \\
    & AdvRAGgen & \textbf{95} & 33  & 66\\

    \midrule
    & HotFlip & \textbf{20} & 0 & 3\\
    \textbf{GTE Large}  & HotFlip (spread out) & \textbf{28} & 2 & 5\\
    \cite{li2023generaltextembeddingsmultistage} & Query as poison & \textbf{58} & 9 & 29 \\
    & AdvRAGgen & \textbf{83} & 15 & 57 \\
    \bottomrule
  \end{tabular}}
\end{table}

\begin{table}[!htbp]
  \caption{\textbf{RAGPart SR  - Intersection based aggregation vs Majority vote based aggregation} in FiQA: This table showcases the ineffectiveness of intersection based aggregation methods as in the case of naive combination. The overly conservative nature of this aggregation makes it impractical.}
  \label{tab:post_sr_intersection_v_voting_}
  \centering
  \resizebox{\textwidth}{!}{\begin{tabular}{c|ccc}
    \toprule
    &  \multicolumn{3}{c}{\textbf{FiQA} \cite{10.1145/3184558.3192301}}  \\
    \midrule
     & \multicolumn{3}{c}{SR $\uparrow$ (\%)} \\
     \midrule
     Retriever & No Defense & \defpart & \defpart  \\
   
     & &( N=5, k=3)  & (N=5, k=3) \\
     & & \textbf{Intersection based aggregation} & \textbf{Majority vote based aggregation } \\
    \midrule
    \textbf{Contriever} & \textbf{58} & 19 & 36 \\
    \cite{izacard2022unsuperviseddenseinformationretrieval}& & & \\
    \midrule
    \textbf{ANCE}  & \textbf{33} & 10 & 18\\
    \cite{xiong2020approximatenearestneighbornegative} & & & \\
    \midrule
    \textbf{Multilingual E5}  & \textbf{79} & 32 & 60 \\
    \cite{wang2024multilinguale5textembeddings} & & & \\ 
    \midrule
    \textbf{GTE Large}  & \textbf{64} & 25 & 43 \\
    \cite{li2023generaltextembeddingsmultistage} & & & \\
    \bottomrule
  \end{tabular}}
\end{table}

\begin{table}[!htbp]
  \caption{\textbf{RAGPart ASR - Intersection based aggregation vs voting based aggregation} in FiQA: Under \defpart due to the added robustness in embedding space under both intersection and majority vote based aggregation \defpart was able defend effectively. But due to the impractical nature of the aggregation in utility preservation we chose majority voting as the ideal aggregation method.}
  \label{tab:post_asr_intersection_v_voting}
  \centering
  \resizebox{\textwidth}{!}{\begin{tabular}{c|c|ccc}
    \toprule
    & & \multicolumn{3}{c}{\textbf{FiQA} \cite{10.1145/3184558.3192301}}  \\
    \midrule
     & & \multicolumn{3}{c}{ASR $\downarrow$ (\%)} \\
     \midrule

      Retriever & Attack & No Defense & \defpart & \defpart  \\
    
     & & & (N=5, k=3)  & (N=5, k=3) \\
      & & & \textbf{Intersection based aggregation} & \textbf{Voting based aggregation } \\

    \midrule
    & HotFlip & \textbf{95} & 5  & 1 \\
    \textbf{Contriever} & HotFlip (spread out) & \textbf{92} & 2  & 6\\
    \cite{izacard2022unsuperviseddenseinformationretrieval}& Query as poison& \textbf{85} & 1 & 1\\
    & AdvRAGgen & \textbf{94} & 3 & 11 \\
    \midrule

    & HotFlip & \textbf{78} & 0 & 0 \\
    \textbf{ANCE}  & HotFlip (spread out) & \textbf{67} &  0 & 1\\
    \cite{xiong2020approximatenearestneighbornegative} & Query as poison & \textbf{65} & 0  & 0\\
   & AdvRAGgen & \textbf{86} & 1 & 4 \\
    \midrule

    & HotFlip & \textbf{21} & 0 & 0\\
    \textbf{Multilingual E5}  & HotFlip (spread out) & \textbf{26} & 0 & 4 \\
    \cite{wang2024multilinguale5textembeddings} & Query as poison & \textbf{45} & 0 & 1 \\
    & AdvRAGgen & \textbf{95} & 6  & 14\\

    \midrule
    & HotFlip & \textbf{20} & 0 & 0\\
    \textbf{GTE Large}  & HotFlip (spread out) & \textbf{28} & 2 & 5\\
    \cite{li2023generaltextembeddingsmultistage} & Query as poison & \textbf{58} & 0 & 0\\
    & AdvRAGgen & \textbf{83} & 2 & 6 \\
    \bottomrule
  \end{tabular}}
\end{table}

\end{document}